\documentclass[10pt,aps,floatfix,citeautoscript,superscriptaddress,prb,
 reprint,      
 amsmath,amssymb,longbibliography
]{revtex4-2}

\usepackage{graphicx}
\usepackage{amsmath, amssymb}
\usepackage{hyperref}
\usepackage{float}
\usepackage[most]{tcolorbox}
\usepackage{xcolor}
\usepackage{setspace}
\usepackage{enumitem}
\usepackage{subcaption}
\usepackage{comment}
\usepackage{tikz}
\usepackage{float}
\usepackage{amsthm}
\theoremstyle{definition}
\newtheorem{definition}{Definition}
\usepackage{enumitem}
\usetikzlibrary{calc}
\usepackage{bbm}
\usepackage{amsfonts}
\usepackage{simpler-wick}
\usepackage[compat=1.1.0]{tikz-feynman}
\usetikzlibrary{patterns}
\usepackage{array}
\usetikzlibrary{positioning,arrows}
\usetikzlibrary{decorations.pathmorphing}
\usetikzlibrary{decorations.markings}
\numberwithin{equation}{section}

\newcommand{\sectionn}{\section}

\begin{document}
\newcommand{\drawHexagon}[2]{
  \begin{scope}[xshift=#1 cm]
    \foreach \i/\angle in {1/30, 2/90, 3/150, 4/210, 5/270, 6/330} {
      \coordinate (P\i) at ({cos(\angle)}, {sin(\angle)});
    }

    \foreach \i in {1,...,6} {
      \pgfmathtruncatemacro{\j}{mod(\i,6)+1}
      \draw (P\i) -- (P\j);
    }

    \node at ($(P1)+(0.3,0.1)$) {\( i_1 \)};
    \node at ($(P2)+(0,0.3)$)   {\( i_2 \)};
    \node at ($(P3)+(-0.3,0.1)$) {\( i_3 \)};
    \node at ($(P4)+(-0.3,-0.1)$) {\( i_4 \)};
    \node at ($(P5)+(0,-0.3)$)   {\( i_5 \)};
    \node at ($(P6)+(0.3,-0.1)$) {\( i_6 \)};

    \foreach \i in {1,...,6} {
      \pgfmathtruncatemacro{\j}{mod(\i,6)+1}
      \path let \p1 = (P\i), \p2 = (P\j) in
        coordinate (M\i) at ($(\p1)!.5!(\p2)$);
    }

    \ifnum#2=1
      \draw (M1) -- (M4);
      \draw (M2) -- (M5);
      \draw (M3) -- (M6);
    \fi
    \ifnum#2=2
      \draw (M1) -- (M5);
      \draw (M2) -- (M6);
      \draw (M3) -- (M4);
    \fi
    \ifnum#2=3
      \draw (M1) -- (M6);
      \draw (M3) -- (M4);
      \draw (M2) -- (M5);
    \fi
        \ifnum#2=4
      \draw (M1) -- (M2);
      \draw (M3) -- (M4);
      \draw (M5) -- (M6);
    \fi
  \end{scope}
}
\newcommand{\drawSquare}[2]{
  \begin{scope}[xshift=#1 cm]
    \coordinate (P1) at (1, 1);   
    \coordinate (P2) at (-1, 1);  
    \coordinate (P3) at (-1, -1); 
    \coordinate (P4) at (1, -1);  

    \foreach \i in {1,...,4} {
      \pgfmathtruncatemacro{\j}{mod(\i,4)+1}
      \draw (P\i) -- (P\j);
    }

    \node at ($(P1)+(0.2,0.2)$) {\( i_1 \)};
    \node at ($(P2)+(-0.2,0.2)$) {\( i_2 \)};
    \node at ($(P3)+(-0.2,-0.2)$) {\( i_3 \)};
    \node at ($(P4)+(0.2,-0.2)$) {\( i_4 \)};

    \foreach \i in {1,...,4} {
      \pgfmathtruncatemacro{\j}{mod(\i,4)+1}
      \path let \p1 = (P\i), \p2 = (P\j) in
        coordinate (M\i) at ($(\p1)!.5!(\p2)$);
    }

    \ifnum#2=1
      \draw (M1) -- (M3);
      \draw (M2) -- (M4);
    \fi
    \ifnum#2=2
      \draw (M1) -- (M2);
      \draw (M3) -- (M4);
    \fi
    \ifnum#2=3
      \draw (M1) -- (M4);
      \draw (M2) -- (M3);
    \fi
    \ifnum#1=1
      \fill (0,0) circle (4pt);
    \fi
  \end{scope}
}
\title{Random Matrix Spectra from Boltzmann-Weighted Lattice Ensembles}
\author{Yaprak Önder}\email{yaprakonder@g.harvard.edu}
    \affiliation{Department of Physics, Harvard University, Cambridge MA, USA, 02139}
    \affiliation{Max Planck Institute for the Physics of Complex Systems, 01187 Dresden, Germany}
    
\author{Abbas Ali Saberi}
\email{asaberi@constructor.university}
\affiliation{School of Science, Constructor University, Campus Ring 1, 28759 Bremen,
Germany}
\affiliation{Max Planck Institute for the Physics of Complex Systems, 01187 Dresden, Germany}
 
\author{Roderich Moessner}
        \affiliation{Max Planck Institute for the Physics of Complex Systems, 01187 Dresden, Germany}

\date{\today}

\begin{abstract}
We introduce a random matrix framework for studying statistical-mechanical lattice systems through spectral observables. Equilibrium configurations sampled from a Boltzmann measure are mapped to matrix ensembles whose covariance structure is inherited from the spatial correlations of the underlying model. This construction maps real-space correlation functions to a momentum-space variance profile, providing a direct bridge between statistical-mechanical correlations and correlated random matrix ensembles. We derive this variance profile in finite-correlation-length and critical regimes, and compute spectral moments within a Wick-contraction expansion. A complementary self-consistent description of the bulk density is developed using the resolvent formalism. These analytical methods are benchmarked against Monte Carlo data for the two-dimensional Ising model and three-dimensional Edwards--Anderson spin glasses. In both cases, the spectra evolve from the semicircle law at high temperature to model-dependent critical forms reflecting the structure of correlations. The framework, therefore, provides a quantitative spectral route to probing collective behavior in ordered and disordered statistical systems, while also defining a class of physically motivated correlated random matrix ensembles.
\end{abstract}
\maketitle

\sectionn{Introduction}
A central goal of statistical physics is to identify collective variables that capture macroscopic emergent behavior without tracking every microscopic degree of freedom. The standard form of this program proceeds through renormalization-group and field-theoretic arguments, supported by large-scale Monte Carlo calculations, and has been remarkably successful in explaining universal collective behavior in systems such as magnets, fluids, and polymers~\cite{WilsonKogut1974,Fisher1998,HohenbergHalperin1977,LandauBinder2021}. The current frontier, however, is progressing towards systems for which the usual language of phases, order parameters, and symmetry breaking does not furnish a settled description. Finite-dimensional spin glasses~\cite{Altieri_2024}, amorphous solids~\cite{BerthierReichman2023}, and active matter~\cite{Shaebani_2020} are prominent examples in which the relevant coarse-grained variables remain unclear or are strongly dependent on the proposed effective theory. Studies of such systems are therefore often targeted towards addressing a particular phenomenological description.

This motivates the search for observables that remain sensitive to collective structure independent of any single candidate description. Modern approaches in this direction include machine-learning methods for phase identification~\cite{CarrasquillaMelko2017} and information-theoretic probes based on conditional mutual information~\cite{ZhangGopalakrishnan2025}. A complementary possibility comes from spectral methods: in statistics, principal component and low-rank approximation techniques are routinely used to reduce high-dimensional data while preserving its relevant correlation structure~\cite{Greenacre2022PCA,TroppWebber2023RandomizedLowRank}. This motivates a similar strategy in statistical physics: can spectral observables provide an informative description of physical systems without relying on model-specific probes? Random matrix theory, which studies the statistical properties of eigenvalue spectra, provides a natural setting in which to address this question.

In previous work, random matrices constructed from the classical Ising model~\cite{Saberi2024} and the three-dimensional Edwards-Anderson spin glass~\cite{sdbx-wx5t} were shown to exhibit distinctive temperature evolutions in their bulk eigenvalue spectra. In particular, their eigenvalue distribution functions at criticality were well-fitted to a known analytical form. These results provide initial evidence that eigenvalue spectra constructed from statistical-physics ensembles can serve as computationally efficient probes of universal collective behavior. This motivates two directions for further study. First, analytical methods from random matrix theory appear to provide a largely unexplored route to extracting information about statistical-mechanics models themselves. Second, the random matrix ensembles induced by statistical-mechanics models inherit a highly structured correlation profile from the underlying many-body system, making them interesting random matrix ensembles for analytical study. Related physically motivated random-matrix ensembles have also been developed for disordered solids and elastic networks~\cite{cicuta2018unifying}. More broadly, this construction raises a conceptual question that has not been systematically explored: given the central role of universality in both statistical physics and random matrix theory, can these two notions of universality be understood within a common framework? The goal of the present work is to develop this connection further and to clarify how spectral observables encode phase structure, criticality, and universality. 

Here, we develop a theoretical framework for random matrix ensembles generated from statistical physics systems. These ensembles are defined by sampling matrix degrees of freedom directly from equilibrium configurations of lattice models, so that the random matrix measure is induced by the Boltzmann distribution of the underlying Hamiltonian. Our primary object is the temperature evolution of the eigenvalue spectrum. We develop methods to relate the shape and moments of the spectrum to the correlation structure of the statistical-mechanical model and describe how scaling behavior arises as the system approaches criticality. Previous work introduced random matrices constructed from the
two-dimensional Ising model and the three-dimensional
Edwards--Anderson spin glass and studied their temperature-dependent
bulk spectra. The present work addresses how the spatial correlation
structure of the underlying Boltzmann ensemble determines the variance
profile, spectral moments, and bulk density of the induced random
matrix ensemble.

For statistical physics systems with power-law correlations, a useful point of comparison is provided by a recent study of long-range correlated random matrices, in which the matrix entries were generated by sign-thresholding Gaussian random fields with algebraic correlations $C(r)\sim r^{-2H}$~\cite{saberi2026longrangecorrelatedrandommatrices}. In these ensembles, long-range correlations generate a hierarchy of moment thresholds and spectral regimes: for $0<H<1/2$ the limiting density has algebraic tails, at $H=1/2$ the support remains unbounded although every fixed moment is finite, and for $H>1/2$ the support is bounded, approaching the semicircle as $H\to\infty$. The long-distance exponent controls the moment thresholds and asymptotic tail, whereas the detailed central density and finite moments depend on the full covariance. The present work extends this idea from externally prescribed power-law fields to matrix ensembles induced directly by Boltzmann-weighted lattice systems at criticality. In this setting, the variance profile is no longer imposed by hand, but is inherited from the physical correlation function of the underlying statistical-mechanical model.

The paper is organized as follows. In Section~\ref{sec:definition}, we define matrix ensembles generated from equilibrium lattice configurations and derive the general correlation structure inherited from the underlying statistical-mechanical system. In Section~\ref{sec:rmtanalytical}, we calculate the effective variance profiles that arise away from criticality and at criticality, discuss their normalization, and identify the features that govern the spectrum in the different temperature regimes. Section~\ref{sec:momentmethod} then introduces a moment-based analysis of the eigenvalue distribution, while Section~\ref{sec:resolvent} develops a complementary resolvent approach based on the Vector Dyson Equation to obtain a self-consistent description of the bulk spectral density. Throughout the paper, the analytical results are compared with Monte Carlo data for the 2D Ising model and the 3D Edwards--Anderson spin glass. We conclude with extensions of the framework and directions for future work.

\section{Definition and Scope\label{sec:definition}}
In this study, we restrict our attention to lattice models with local degrees of freedom defined on a discrete $d$-dimensional hypercube, with sites labeled $\mathbf{x}=(x_1,x_2...,x_d)$, unit lattice spacing, and linear size $L$ in each direction. Periodic boundary conditions are imposed in all dimensions to obtain a translationally-invariant system, and the system is assumed to respect the discrete rotational and reflection symmetries of the hypercubic lattice. These assumptions ensure that the matrix ensemble inherits the translation and point-group symmetries of the system, which simplify its analytical study. We discuss possible generalizations to different geometries in the conclusion. 

The matrix entries are sampled from a scalar parameter field $q(\mathbf{x})$ defined on the same lattice. The choice of $q(\mathbf{x})$ therefore determines how the underlying physical correlations are transferred to the random matrix ensemble. For instance, in our previous studies, we chose the spin value at each site for Ising model random matrices~\cite{Saberi2024}, and the Parisi replica overlap parameter for Edwards-Anderson spin glasses~\cite{sdbx-wx5t}. 

For models in dimension $d> 2$, we reduce the dimensionality of the model to $d=2$ to enable an eigenvalue decomposition. This is done straightforwardly by studying an arbitrary 2D slice of our parameter field, e.g. $q_{xy}:=q(x,y,0,\dots,0)$, where all other indices are set to zero without loss of generality due to translational invariance. We define a matrix on this 2D slice by using the spatial coordinates of the parameter field as matrix indices, inserting the value of the parameter field in its appropriate position: 
\begin{equation}
    \mathcal{M'}_{xy} = q({x,y}).
\end{equation}
With this choice, spatial correlations between degrees of freedom correspond directly to correlations between matrix elements. Translation invariance imposes the following generic correlation profile: 
\begin{equation}
 \langle q(\mathbf{x})   q(\mathbf{y})\rangle = G(\mathbf{x}-\mathbf{y}),
\end{equation}
where the subtraction is performed on the discrete torus $(\mathbb Z/L\mathbb Z)^d$, i.e., all differences are taken mod $L$.

In general, the Hamiltonian of the system of interest will be a joint functional $H[q,\mathbf{q'}]$ of the parameter field $q(\mathbf{x})$, and additional degrees of freedom $\{\mathbf{q'}\}$. The probability distribution of the matrices $\mathcal{M'}$ is then obtained by integrating the Boltzmann measure over the degrees of freedom not explicitly retained in $\mathcal{M}(q)$:
\begin{equation}
    P(\mathcal{M}) \propto \int_{\Lambda_{q}} dq\int_{\Lambda_{q'}} d\mathbf{q'} e^{-\beta H(q,\mathbf{q'})}\delta(\mathcal{M}-\mathcal{M}[\mathbf{q}]).
\end{equation}
The resulting random matrix ensemble is thus defined by a measure induced by the Boltzmann distribution of the Hamiltonian. 

For the case of disordered systems, one can construct a matrix ensemble from independent replicas drawn from the Gibbs measure at fixed disorder and then average over disorder realizations. For example, the matrix $\mathcal{M}(\mathbf{s},\mathbf{s'})$ can be constructed from the two-replica overlap:
\begin{align}
P(\mathcal M)
&\propto \int d\mathbf J\, P(\mathbf J)\, P(\mathcal M\mid \mathbf J), \\
&\propto \int d\mathbf J\, P(\mathbf J)
   \int_{\Lambda_s} d\mathbf s \int_{\Lambda_{s'}} d\mathbf s'\,
   e^{-\beta H_{\mathbf J}(\mathbf s)} e^{-\beta H_{\mathbf J}(\mathbf s')}\notag\\
&\qquad\qquad \qquad \qquad \qquad \times 
   \delta\!\big(\mathcal M(\mathbf s,\mathbf s')-\mathcal M\big).
\end{align}
where $\mathbf{s,s'}$ denote two independent configurations sampled from the Gibbs measure of the Hamiltonian $H_\mathbf{J}$ with disorder realization $\mathbf{J}$. 

We now give two examples of how this procedure may be performed. In this paper, we will explicitly compare our analytical predictions to Monte Carlo results for spin-$1/2$ models with nearest-neighbor interactions: 
\begin{equation}
    H = -\sum_{\langle \mathbf{x,y}\rangle } J_{\mathbf{x,y}}S_\mathbf{x}S_\mathbf{y},
\end{equation}
where we consider two choices of couplings:
\begin{enumerate}
    \item 2D ferromagnetic Ising model, for which $J_{\mathbf{x,y}}=1$ on every nearest-neighbor bond.
    \item 3D Edwards-Anderson spin glass, for which the couplings $J_{\mathbf{x,y}}$ are quenched random variables drawn independently and identically on each nearest-neighbor bond from a unit Gaussian distribution. 
\end{enumerate}
For the 2D Ising model on a square lattice we use the spin fields as our degrees of freedom. 
\begin{equation}
    \mathcal{M'}_{ij} = S_{ij}.
\end{equation}
For the EA model, the order parameter is typically taken as the overlap between two replicas:
\begin{equation}
    q_{\mathbf{x}} = \begin{cases}
        1 & \text{if } s^{(1)}_{\mathbf{x}} = s^{(2)}_{\mathbf{x}}, \\
        -1 & \text{otherwise},
    \end{cases}
\end{equation}
where the two replicas, denoted as (1) and (2), share the same disorder realization $J_{\mathbf{xy}}$. The Parisi overlap parameter is then defined as $\frac{1}{L^d}\sum_{\mathbf{x\in}s} q_{\mathbf{x}}$. We define $L \times L$ matrices based on the Parisi overlap parameter field $q$ calculated on 2D cross-sections of two replicas of the 3D system 

\begin{equation} 
    \mathcal{M}'_{ij} =
       q_{ij},
\end{equation}
where $q_{ij}$ represents the overlap site evaluated at an arbitrary 2D slice of the 3D spin glass system.

To ensure a real spectrum, and thereby make contact with the bulk of the random matrix theory literature, we Hermitize our matrices and choose the Wigner normalization:
\begin{equation}
    \mathcal{M} = \frac{\mathcal{M'}+\mathcal{M'}^\dagger}{\sqrt{2L}}.\label{eqn:matrixdef}
\end{equation}
The above procedure is equivalent to symmetrization for real parameter fields. With this convention, the raw field has entries of order one, while the Hermitian matrix $\mathcal{M}$ has matrix elements of order $L^{-1/2}$. In the high-temperature limit of uncorrelated local variables, this normalization gives the standard Wigner scaling.
The PDF of the random matrix ensemble induces a joint PDF $P\{\lambda_1,\lambda_2,...\lambda_L\}$ of matrix eigenvalues. In this work, we concentrate on the global eigenvalue scale, studying the \textit{empirical spectral distribution} which describes the macroscopic distribution of all eigenvalues in a given realization. 

The correlation profile induces its 2-dimensional projection onto the unsymmetrized matrix ensemble
\begin{equation}
    \langle \mathcal{M'}_{xy}  \mathcal{M}^{'*}_{zw}\rangle = G(x-z,y-w) ,
\end{equation}
which induces the following form on the symmetrized matrix ensemble:
\begin{equation}
    \langle\mathcal{M}_{xy}  \mathcal{M}^*_{zw}\rangle = \frac{1}{L}(G(x-z,y-w) + G(x-w,y-z))\label{eqn:TIcorrelations},
\end{equation}
where all differences are taken modulo $L$. Matrices satisfying \eqref{eqn:TIcorrelations} were previously considered in~\cite{Pastur1996Eigenvalue,anderson2007lawlargenumbersfiniterange,Ajanki_2016}, in which a discrete Fourier transformation was used to obtain an ensemble with more tractable correlation structure. Applying random matrix theory  techniques to correlated random matrices is not straightforward; however, a mapping to momentum space generates an uncorrelated matrix up to an additional symmetry. The discrete Fourier transformation is generated by the unitary conjugation 
\begin{equation}
    {\hat{\mathcal{M}}} = \mathcal{FMF}^\dagger,
\end{equation}
where the Fourier matrix $\mathcal{F}$ has entries
\begin{equation}
    \mathcal{F}_{{k_x,x}}=\frac{1}{\sqrt{L}}e^{2\pi ik_xx/L}. \label{eqn:deffour}
\end{equation}
We index the Fourier-transformed matrix with integers $k_x,k_y=-L/2+1,...L/2$ such that the Fourier modes $\mathbf{k}=\frac{2\pi}{L}(k_x,k_y)$ lie within the first Brillouin Zone. The resulting correlation profile is uncorrelated up to an additional constraint:
\begin{equation}
    \langle \hat{\mathcal{M}}_{k_xk_y}\hat{\mathcal{M}}_{q_xq_y}\rangle =  (\delta_{k_x,q_y}\delta_{k_y,q_x} + \delta_{k_x,-q_x}\delta_{k_y,-q_y})\mathcal{V}_{k_xk_y}, \label{eqn:matvarcor}
\end{equation}
where the \textit{matrix of variances} $\mathcal{V}_{k_xk_y}$ is related to the real-space correlation structure by
\begin{equation}
\mathcal{V}_{k_xk_y}
:= \frac{1}{L}\sum_{\Delta x\in \mathbb{Z}_L} \sum_{\Delta y\in \mathbb{Z}_L}
e^{\frac{2\pi i}{L} (k_x \Delta x - k_y \Delta y)}
G(\Delta x, \Delta y).
  \label{eqn:matvardef}
\end{equation}
A derivation of \eqref{eqn:matvarcor} is given in Appendix~\ref{sec:corrcalculation}. Note that matrix entries which are uncorrelated under the law \eqref{eqn:matvarcor} are independent only under the additional assumption that they are jointly Gaussian. This is the setting considered in \cite{Ajanki_2016}. In the absence of the second term, which couples Fourier modes of opposite sign, the covariance structure reduces to that of a Wigner-type ensemble; in the jointly Gaussian case, this is precisely a Wigner-type matrix, studied extensively in \cite{ajanki2017universalitygeneralwignertypematrices,Ajanki_2019}. 
 We will later show that, in the $L\rightarrow\infty$ limit, the relevant global properties of the eigenvalue spectrum are determined by the first term in eq.~\eqref{eqn:matvarcor}. 

Note that, when all matrix elements are identically Gaussian distributed, i.e. $\mathcal{V}_{mn}=1/L$ for all indices, the matrix is known as a \textit{Wigner matrix} and its eigenvalue statistics follow the Wigner semicircle distribution independent of the specific distribution of matrix elements~\cite{mehta1991random}. We expect statistical physics random matrices to conform to this universal behavior in the $T\rightarrow\infty$ limit, where thermal fluctuations suppress correlations between matrix elements. 

Away from criticality, statistical-mechanical ensembles are well-approximated at the level of two-point structure by Gaussian fluctuations, and in this regime the random matrix ensembles considered here are correspondingly well described by Wigner-type matrices. The analysis is more involved at criticality, both because of long-range power-law correlations and because higher-order cumulants need not be negligible. In the analytical treatment below, we consider the off-critical and critical regimes separately. 

\section{Defining the Statistical Mechanical Ensembles: Off-Critical and Critical Behavior\label{sec:rmtanalytical}}
Having established the general form of the variance profile for translationally invariant systems, we now characterize the classes of random matrix ensembles that arise from statistical mechanical models. Since relevant matrix correlations are encoded in the matrix of variances $\mathcal{V}$, we define these ensembles through their momentum-space correlation profile. Our goal is to identify how the correlation structure of the underlying lattice system is encoded in $\mathcal{V}$, and how this structure changes between off-critical and critical regimes. 

We proceed by expressing $\mathcal{V}$ in terms of the momentum-space two-point function of the underlying field. Starting from eq.~\eqref{eqn:matvardef}, we define the Fourier transform of the correlation function as
\begin{equation}
    G(\mathbf{k}) = \frac{1}{L^d}\sum_{\mathbf{x}\in\mathbb{Z}^d_L} e^{i\mathbf{k\cdot x}} G(\mathbf{x}).
\end{equation}
Using standard Fourier identities on the discrete torus, the variance matrix can be written as a projection of the full $d$-dimensional momentum-space correlator onto the two-dimensional subspace defining the matrix ensemble:
\begin{equation}
\mathcal{V}_{k_1,k_2} \sim \sum_{q_1,\dots,q_{d-2}} G(\mathbf{k},\mathbf{q}).
\end{equation}

Here, $\mathbf{k}=(k_1,k_2)$ denotes the Fourier modes of the matrix indices, while $\mathbf{q}$ represents the remaining $(d-2)$ momentum components spanning the full Brillouin zone of the underlying lattice. In the continuum limit, we replace discrete sums by integrals according to
\begin{equation}
    \sum_{p=-L/2}^{L/2} \rightarrow \int_{-\pi}^{\pi} \frac{L}{2 \pi} \, dk.
\end{equation}
Unless otherwise stated, all integrals are taken over the first Brillouin zone $\Lambda^d=[-\pi,\pi]^d$. 
\subsection{$T> T_c:$ Short-Range Interaction-Correlated Random Matrices}
Away from criticality, the standard Ornstein-Zernike relation characterizes the Fourier-space decay form~\cite{goldenfeld,OrnsteinZernike1914}
\begin{equation}
     G(\mathbf{k}) :=\frac{1}{L^d}\sum_{\mathbf{r\in}\mathbb{Z}_L^d}e^{i\mathbf{k\cdot r}}G(\mathbf{r}) \sim \frac{1}{\mathbf{|k|}^2+\frac{1}{\xi^2}}\label{eqn:ornsteinzernike}.
\end{equation}
The leading decay form of two-point connected correlation functions is characterized by exponential decay with finite correlation length $\xi$:
\begin{equation}
    G(\mathbf{r}):=\langle S_{\mathbf{x}}S_{\mathbf{x+r}}\rangle -  \Theta(T_c-T)\langle |S_{\mathbf{x}}|\rangle^2 \sim e^{-r/\xi}.
\end{equation} 
The matrix of variances of a given off-critical ensemble can then be calculated as follows:
\begin{align}
\mathcal{V}_{k_x,k_y}
& =  A(L,\xi,d) \hat G_d(\mathbf{k}), \\
\hat G_d(\mathbf{k}) & = 
\int \frac{d^{d-2}\mathbf{q}}{(2\pi)^{d-2}} \frac{1}{\mathbf{k}^2+\mathbf{q^2}+\frac
{1}{\xi^2}}, \label{eqn:genvar}
\end{align}
where $A(L,\xi,d)$ is the normalization factor. This reduces to the expected Ornstein-Zernike form~\eqref{eqn:ornsteinzernike} for $d=2$ and has a closed-form solution for $d=3$
\begin{equation}
    \mathcal{V}(\mathbf{k}) \sim \frac{A(L,\xi,d) }{\sqrt{\mathbf{|k|^2}+ \frac{1}{\xi^2}}}\arctan\left(\frac{\pi}{\sqrt{|\mathbf{k}|^2+\frac{1}{\xi^2}}}\right)\text{ for d = 3}\label{eqn:ornvar3d}.
\end{equation}
Note that, for $d>2$ in the small correlation length ($\xi\ll 1$) limit, the Ornstein-Zernike form is a good approximation to the momentum space correlation structure. 

Additionally, off-critical statistical physics ensembles are well-approximated by Gaussian field theories~\cite{kardar2007statistical}. Therefore, matrix elements are independent except for correlations of the form~\eqref{eqn:matvardef}. 
Small $|\mathbf{k}|$ corrections to the Ornstein-Zernike behavior involving higher-order terms in the denominator can be computed on a system-by-system basis (for an example on the 2D Ising model, see~\cite{Mart_n_Mayor_2002}). Within the Gaussian approximation, we provide the following definition of a random matrix ensemble describing a statistical physics system away from criticality.

\begin{definition}[Short-Range Interaction-Correlated Random Matrix]
A short-range interaction correlated random matrix (\textbf{SIC($L,\xi,d$)}) with correlation length $\xi$ and system dimension $d$ is a $L\times L$ Hermitian matrix with jointly Gaussian-distributed matrix elements and the following momentum-space covariance profile:
\begin{align}
    \langle \hat{\mathcal{M}}_{k_xk_y}\hat{\mathcal{M}}_{q_xq_y}\rangle  =  (\delta_{k_x,q_y}\delta_{k_y,q_x} + \delta_{k_x,-q_x}\delta_{k_y,-q_y})\mathcal{V}_{k_xk_y}, \\
    \mathcal{V}_{k_x,k_y}  = A(L,\xi,d)\int \frac{d^{d-2}\mathbf{q}}{(2\pi)^{d-2}} \frac{1}{(k_x^2+k_y^2)+\mathbf{q^2}+\frac
{1}{\xi^2}}.
\end{align}
\end{definition}
Note that we expect \textbf{SIC($L,\xi,d$)}$\rightarrow$ GOE(L) in the $\xi \rightarrow0$ ($T\rightarrow \infty$) limit. 

\subsection{$T=T_c$ Regime: Long-Range Interaction Correlated Random Matrices\label{sec:longrange}}
At criticality ($T = T_c$), the momentum-space correlator develops an infrared divergence due to the absence of a finite correlation length. In finite systems, this divergence is regularized by the minimal momentum scale set by the system size. Accordingly, we introduce a lower-momentum cutoff
\begin{equation}
\Lambda' = \left\{ \mathbf{k} \;\middle|\; \frac{2\pi}{L} < |\mathbf{k}| < 2\pi \right\},
\end{equation}
which excludes the zero mode. The $\mathbf{k}=0$ contribution will be treated separately, as it corresponds to the global order parameter and leads to a macroscopic contribution to the variance matrix $\mathcal{V}$.

Critical systems are characterized by power-law decay profiles
\begin{equation}
    G(\mathbf{r}) \sim \frac{1}{|\mathbf{r}|^{d-2+\eta}}\label{eqn:powerlawreal}.
\end{equation}
where $\eta$ is the anomalous dimension~\cite{goldenfeld}. This induces the following form on the matrix of variances: 
\begin{align}
   \mathcal{V}_{k_x,k_y}  = \mathcal{A}(L,\eta,d)\int \frac{d^{d-2}\mathbf{q}}{(2\pi)^{d-2}} \frac{1}{|(k_x^2+k_y^2)+\mathbf{q^2}|^{1-\eta/2}} \label{eqn:longrangeintegral}.
\end{align}
For $d<4-\eta $ a simple scaling analysis of the integral in eq.~\eqref{eqn:longrangeintegral} leads to a power law form 
\begin{equation}
    \mathcal{V}(\mathbf{k}) \sim  \frac{A(L,\eta,d)}{|\mathbf{k}|^{4-d-\eta}} \text{ for } d < 4-\eta. \label{eqn:powerlawk}
\end{equation}
For $d >4-\eta$, the variance profile crosses over from a constant to an inverse square decay law at a characteristic $O(1)$ momentum scale. In the marginal case $d=4-\eta$, the low-momentum form of the correlators instead acquires a logarithmic form. This behavior can be summarized as
\begin{align}
    \mathcal{V}(\mathbf{k}) \sim 
      \begin{cases}
      \text{const.} \text{ for } |\mathbf{k}| \ll k_c,  \; d>4-\eta,\\
        -\ln(\mathbf{|k|})\text{ for } |\mathbf{k}| \ll k_c, \; d = 4-\eta,   \\
        \frac{1}{|\mathbf{k}|^2} \text{ for } |\mathbf{k}| \gg k_c,\; d\geq 4-\eta .
    \end{cases} \label{eqn:IRdiv}
\end{align}
where $k_c$ is an $O(1)$ characteristic system-dependent momentum scale. 

When the anomalous dimension $\eta = 0$, as is appropriate for a critical theory at the Gaussian fixed point (e.g., the Ising model in $d \geq 4$), the propagator at criticality can be obtained as the $\xi \rightarrow\infty$ limit of the propagator in eq.~\eqref{eqn:genvar}. 

For finite T, correlations between matrix elements lead to a peak developing in the $\mathbf{k}=0$ element of the matrix $\mathcal{V}$; thus, it must be treated separately from the rest of the entries. The width of this peak decreases as the temperature decreases, with a sharp peak developing for $T\leq T_c$. Using eq.~\eqref{eqn:matvardef}, we can relate this peak to the order parameter $q=L^{-2}\sum_{xy}q_{xy}$ of the matrix ensemble:
\begin{align}
    \mathcal{V}_{00} & = \frac{1}{L} \sum_{\Delta x}^L\sum_{\Delta y}^L G(\Delta x, \Delta y),  \\
    & = \frac{1}{L}\langle \sum_{\Delta x}^L\sum_{\Delta y}^L q(0) q(\Delta x,\Delta y) \rangle, \\
    & = L  \langle q^2\rangle. \label{eqn:V00}
\end{align} 
Thus, if $\langle q^2\rangle\sim O(1)$, as is typical for $T<T_c$, the $\mathbf{k}=0$ mode has a macroscopic occupation, and its behavior separates from the rest of the Fourier modes. We can relate this to the susceptibility~\cite{goldenfeld}:
\begin{equation}
    \chi=\beta L^{d}\langle q^2\rangle = \beta L^{d-1}\mathcal{V}_{00} \label{eqn:chiV00}.
\end{equation}
The susceptibility has a finite-size scaling form $\chi = L^{2-\eta}f(tL^{1/\nu})$. Hence, $\mathcal{V}_{00}\sim L^{3-d-\eta}f(tL^{1/\nu})$. 

We now formalize the discussion in this section by introducing a class of random matrix ensembles exhibiting power-law correlations:
\begin{definition}[Long-Range Interaction-Correlated Random Matrix]
A long-range interaction correlated random matrix (\textbf{LIC$(L,\eta,d)$}) is defined by the following momentum-space covariance profile:
\begin{align}
    \langle \hat{\mathcal{M}}_{k_xk_y}\hat{\mathcal{M}}_{q_xq_y}\rangle  =  (\delta_{k_x,q_y}\delta_{k_y,q_x} + \delta_{k_x,-q_x}\delta_{k_y,-q_y})\mathcal{V}_{k_x,k_y}, \\
    \mathcal{V}_{k_x,k_y}  = \mathcal{A}(L,\eta,d)\int \frac{d^{d-2}\mathbf{q}}{(2\pi)^{d-2}} \frac{1}{|(k_x^2+k_y^2)+\mathbf{q^2}|^{1-\eta/2}}.
\end{align}
\end{definition}
This definition is directly related to the long-range correlated random matrix ensembles studied in~\cite{saberi2026longrangecorrelatedrandommatrices} for $d=2$. In that work, the matrix elements are obtained by sign-thresholding a Gaussian field whose covariance decays as $r^{-2H}$; thresholding modifies the short-distance covariance but preserves this long-distance exponent. In the present notation, a two-dimensional power-law ensemble with $G(r)\sim r^{-\eta}$ therefore corresponds to the identification $\eta=2H$. Thus, the covariance profile
\begin{equation}
    \mathcal V(\mathbf k)\sim |\mathbf k|^{-(2-\eta)}
\end{equation}
for $d=2$ reproduces the infrared structure of the long-range correlated ensemble. The important difference is conceptual: Ref.~\cite{saberi2026longrangecorrelatedrandommatrices} treats $H$ as an externally tunable correlation exponent and shows that the long-distance covariance controls the moment thresholds and asymptotic spectral tail, while the full covariance controls the detailed limiting density and finite shape measures. It therefore provides a controlled benchmark for the covariance-to-spectrum mechanism developed here, where $\eta$, $\xi$, and the order-parameter mode are determined by the Boltzmann ensemble itself.

\subsection{Normalizing the Variance Profile \label{sec:norm}}
In this section we describe a normalization identity that holds when local degrees of freedom in the statistical mechanical ensemble have a definitive magnitude. Without loss of generality, we consider systems for which the local degrees of freedom, before the symmetrization procedure, have unit magnitude, i.e. $|\mathcal{M'}_{xy}|=1~\forall~(x,y)$. With the normalization convention of eq.~\eqref{eqn:matrixdef}, the variance profile is normalized through eq.~\eqref{eqn:matvardef}. Evaluating the element-wise sum $\sum_{k_xk_y}\mathcal{V}_{k_x,k_y}$ gives
\begin{align}
  \sum_{k_x,k_y}^L\mathcal{V}_{k_x,k_y}
  &= \frac{1}{L}\sum_{\Delta x,\Delta y\in\mathbb{Z}_L}
  G(\Delta x,\Delta y)
  \sum_{k_x,k_y}^L
  e^{\frac{2\pi i}{L}(k_x\Delta x-k_y\Delta y)},\nonumber\\
  &= L\,G(0,0)=L. \label{eqn:conservation}
\end{align}
\begin{figure*}[htbp]
  \centering
  \begin{subfigure}[t]{0.32\textwidth}
    \centering
    \includegraphics[width=\textwidth]{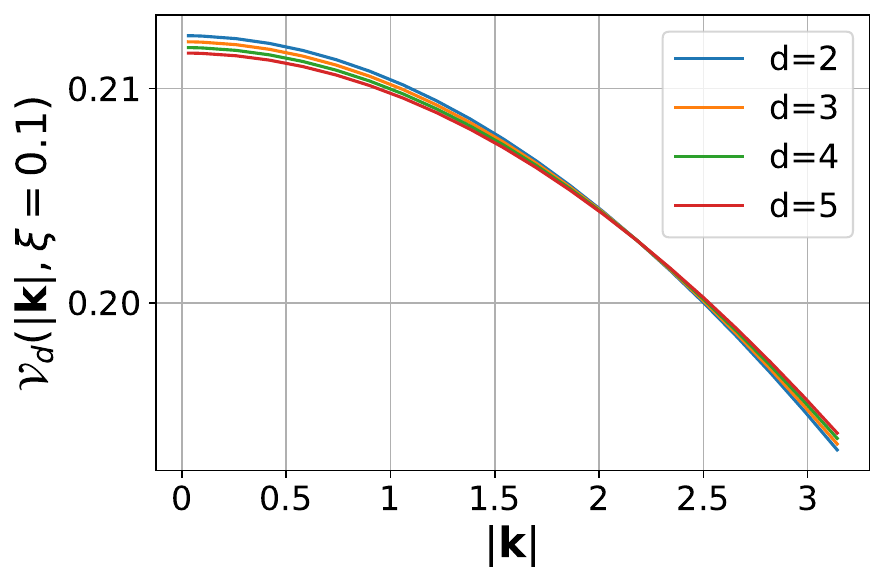}
    \label{fig:sub1}
  \end{subfigure}
  \begin{subfigure}[t]{0.32\textwidth}
    \centering
    \includegraphics[width=\textwidth]{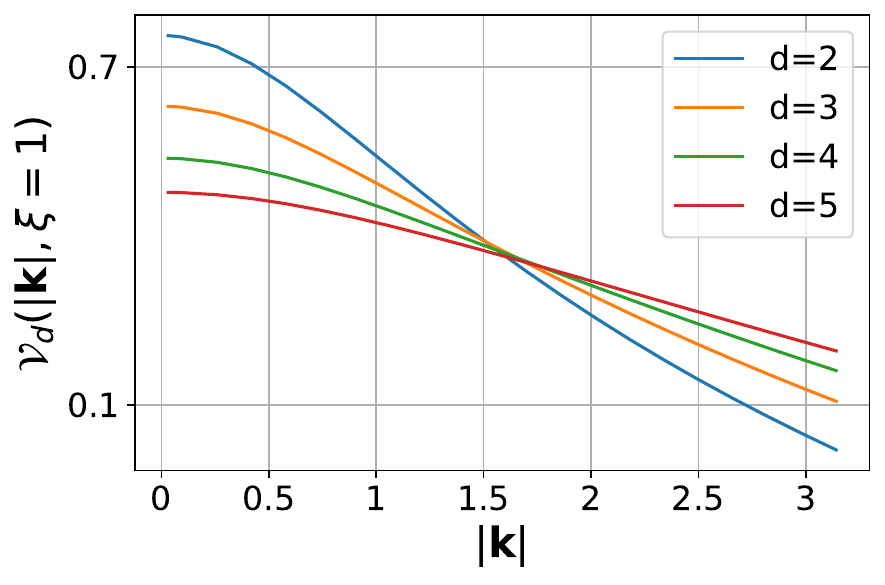}
  \end{subfigure}
  \begin{subfigure}[t]{0.32\textwidth}
    \centering
    \includegraphics[width=\textwidth]{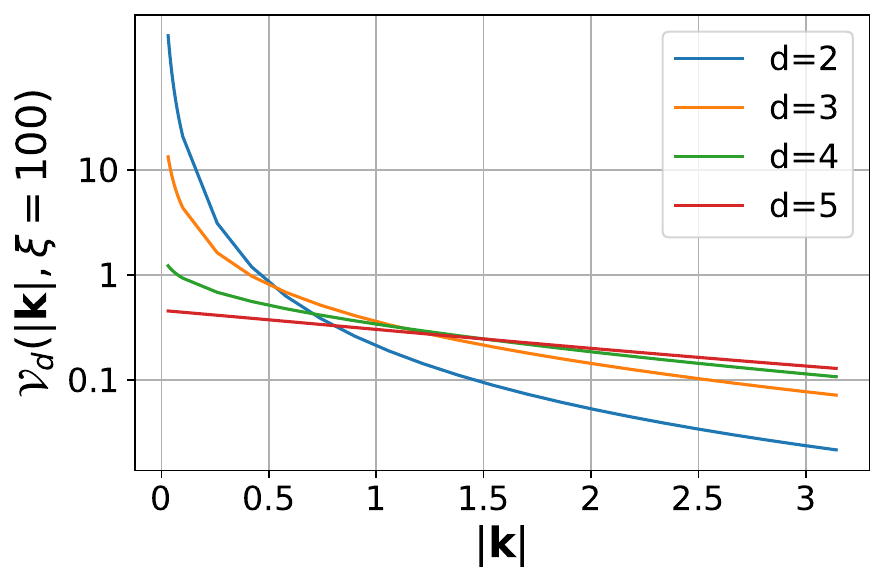}
  \end{subfigure}
  \caption{Projected propagator $\mathcal V_d(k,\xi)$ as a function of
the momentum magnitude $k=|\mathbf{k}|$ for $d=2,3,4,5$ and several values
of $\xi$.}
  \label{fig:propagatorxi}
\end{figure*}
In the last step we used the discrete Fourier orthogonality relation and the unit-magnitude condition $G(0,0)=\langle |q_{xy}|^2\rangle=1$. Thus we have a normalization constraint that is independent of the temperature.

We define the following integral, which will be used to set the normalization in our calculations:
\begin{align}
 \int_{\Lambda^d}\frac{d^d \mathbf{k}}{(2\pi)^d} \frac{1}{\mathbf{k^2} + \frac{1}{\xi^2}} := I_d(\xi)
\end{align}
For $T>T_c$, the conservation law in eq.~\eqref{eqn:conservation} holds, 
\begin{align}
   L^2\int_{\Lambda^2} \frac{d^2 \mathbf{k}}{(2\pi)^2}\mathcal{V}_d(\mathbf{ k})& =L^2 \int_{\Lambda^d}\frac{d^d \mathbf{k}}{(2\pi)^d} \frac{A(L,\xi,d)}{\mathbf{k^2} + \frac{1}{\xi^2}},\nonumber \\
   & = L^2 A(L,\xi,d)I_d(\xi) = L.
   \label{eqn:Idc}
\end{align} 
Setting $\sum_{k_x,k_y}\mathcal{V}_{k_x,k_y}=L$, we obtain $A(L,\xi,d) = (LI_d(\xi))^{-1}$
\begin{equation}
    \mathcal{V}_d(\mathbf{k}) =  \frac{\hat{G}_d(\mathbf{k},\xi)}{L I_d(\xi) } \label{eqn:propagatorhot},
\end{equation}
where $\hat{G}_d$ is defined in eq.~\eqref{eqn:genvar}. 
The behavior of $\mathcal{V}(\mathbf{k})$ in dimensions $d=\{2,3,4,5\}$ for various values of $\xi$ is plotted in Fig.~\ref{fig:propagatorxi}. In the limit $\xi \rightarrow \infty$, the propagator has an IR divergence for $d \leq 4$ and is finite for $d > 4$, which matches eq. \eqref{eqn:IRdiv}.   

For the $T=T_c$ regime, the correlations take on a power-law form as indicated in eq.~\eqref{eqn:powerlawk}. We will assume that $d < 4-\eta$ to evaluate the effect of IR divergences at criticality: 
\begin{align}
 L^2\int_{\Lambda'^2}\frac{d^2\mathbf{k}}{(2\pi)^2}& \frac{1}{|\mathbf{k}|^{4-d-\eta}}   := L^2I_d(\eta), \\ &  \sim \left(\frac{L}{2\pi}\right)^2\int_{2\pi/L}^\pi dk  \frac{2\pi}{k^{3-d-\eta}}  \sim L^2. 
\end{align}
Here, we have switched to integrating over a circular region to get a sense of the scaling with $L$. To get a decaying form in real-space we require $\eta> 2-d$. Hence, the IR behavior of the integral, which has the scaling $\sim L^{-(d-2+\eta)}$, is always convergent. One caveat here is that the central mode has a macroscopic occupation for $T\leq T_c$ and thus needs to be considered separately. For $T=T_c$, as shown in eq.~\eqref{eqn:chiV00}, the central mode scales as $\mathcal{V}_{00}\sim L^{3-d-\eta}$, which makes a subextensive contribution to the variance sum in the thermodynamic limit. Thus, in this regime 
\begin{equation}
    \mathcal{V}_{mn} =  \frac{1}{L I_d(\eta) } \frac{1}{\mathbf{|k|}^{4-d-\eta}}.
\end{equation}
At criticality, the finite-correlation-length covariance must be
replaced by the power-law variance profile of Sec.~\ref{sec:longrange}, with the
infrared cutoff set by $L$. Below $T_c$, the macroscopic
order-parameter mode is separated from the connected nonzero-mode
fluctuations as follows:
\begin{equation}
   \sum_{k_x,k_y}\mathcal{V}_{k_x,k_y} = \langle q^2\rangle L+\sum_{k_x,k_y\neq (0,0)}\mathcal{V}_{k_x,k_y} = L,
\end{equation}
Thus, the normalization constant of the nonzero-mode propagator for $T<T_c$ is rescaled by a factor of $(1-\langle q^2\rangle)$ compared to the $T>T_c$ case, assuming that the connected correlators decay exponentially with a finite correlation length $\xi$. The covariance framework thus
extends to both regimes, although higher connected cumulants need not
be negligible at criticality.
\begin{figure*}[!t]
    \centering
    \begin{subfigure}[t]{0.15\linewidth}
        \centering
        \begin{tikzpicture}[scale=0.9]
            \drawSquare{0}{1}
        \end{tikzpicture}\\[-0.25em]
        \textbf{(4-i)}
    \end{subfigure}
    \begin{subfigure}[t]{0.15\linewidth}
        \centering
        \begin{tikzpicture}[scale=0.9]
            \drawSquare{0}{2}
        \end{tikzpicture}\\[-0.25em]
        \textbf{(4-ii)}
    \end{subfigure}
        \begin{subfigure}[t]{0.15\linewidth}
        \centering
        \begin{tikzpicture}[scale=0.9]
            \drawHexagon{0}{1}
        \end{tikzpicture}\\[-0.25em]
        \textbf{(6-i)}
    \end{subfigure}
    \begin{subfigure}[t]{0.15\linewidth}
        \centering
        \begin{tikzpicture}[scale=0.9]
            \drawHexagon{0}{2}
        \end{tikzpicture}\\[-0.25em]
        \textbf{(6-ii)}
    \end{subfigure}
    \begin{subfigure}[t]{0.15\linewidth}
        \centering
        \begin{tikzpicture}[scale=0.9]
            \drawHexagon{0}{3}
        \end{tikzpicture}\\[-0.25em]
        \textbf{(6-iii)}
    \end{subfigure}
    \begin{subfigure}[t]{0.15\linewidth}
        \centering
        \begin{tikzpicture}[scale=0.9]
            \drawHexagon{0}{4}
        \end{tikzpicture}\\[-0.25em]
        \textbf{(6-iv)}
    \end{subfigure}
    \vspace{1em}
    \caption{Distinct pairings (up to a symmetry factor) for the 4th and 6th moments of Wigner matrices. We have identified the matrix indices with the vertices and the matrix elements with the edges; pairings are indicated via lines connecting the edges inside the $n$-gon. Non-crossing pairings (\textbf{4-ii},\textbf{6-iii},\textbf{6-iv} in the figure) contribute at $O(1)$ in the large-$L$ limit, while crossing ones are suppressed.}
    \label{fig:pairings}
\end{figure*}
\section{Moment Method\label{sec:momentmethod}} 
In this section, we outline a technique for approximating the moments of the eigenvalue spectrum using the form of the variance profile $\mathcal{V}$. We will restrict our attention to models with inversion symmetry, and thus assume all odd moments vanish. The moments are related to sums over products of matrix elements as follows:
\begin{align}
    \langle\lambda^{2p} \rangle
    &=
    \int d\lambda\, P(\lambda)\lambda^{2p}
    =
    \frac{1}{L}\operatorname{Tr}(\mathcal{M}^{2p}),
    \nonumber\\
    &=
    \frac{1}{L}
    \sum_{i_1,\ldots,i_{2p}=1}^{L}
    \mathcal{M}_{i_1i_2}
    \mathcal{M}_{i_2i_3}
    \cdots
    \mathcal{M}_{i_{2p}i_1}.
    \label{eqn:moments}
\end{align}

To obtain an exact expression for the moments, all nonvanishing cumulants contributing to the sum~\eqref{eqn:moments} would have to be included. Here, we approximate the moment expansion by 
retaining only pairwise contractions of the matrix elements. This corresponds to a Gaussian, or Wick, approximation for the induced random matrix ensemble. Systems described by an interacting, non-Gaussian field theory generally contain higher connected cumulants, which generate additional contributions to the moment expansion. Such contributions can become important near criticality~\cite{kardar2007statistical}. In the critical regime, we calculate the finite-size scaling form of the moments using scaling theory. 

The pairing procedure is illustrated diagrammatically for the fourth and sixth moments in Fig.~\ref{fig:pairings}, in which a catalog of all distinct nonzero contributions (up to a combinatorial factor) is shown. The covariance structure in eq.~\eqref{eqn:matvarcor} imposes constraints in momentum space, effectively reducing the set of allowed contractions. In the thermodynamic limit, the dominant pairing rule simplifies to
\begin{equation}
\langle\mathcal{M}_{k_x,k_y}\mathcal{M}_{q_x,q_y}\rangle 
= \delta_{k_x,q_y}\delta_{k_y,q_x}\,\mathcal{V}_{k_x,k_y}.
\end{equation}
In the large-$L$ limit, the leading $O(1)$ contributions arise from pairings that maximize the number of independent index summations. This selects non-crossing pairings, while crossing pairings are suppressed by powers of $L$. 

For Wigner matrices, which are identically distributed, the weight factor of all pairings is identical. The problem of calculating the $2p$-th moment of the spectrum thus reduces to that of counting the number of distinct non-crossing pairings of $2p$ edges. This is a well-known combinatorial problem, and the answer is given by the Catalan number 
\begin{equation}
    \langle\lambda^{2p} \rangle = \frac{(2p)!}{p!(p+1)!},
    \label{eqn:semicirclemoment}
\end{equation}
which is equal to the $2p$-th moment of a semicircle distribution with radius $R=2$~\cite{Anderson_Guionnet_Zeitouni_2009}. Statistical physics random matrices in the $T\rightarrow\infty$ limit will conform to this behavior. 

At finite temperatures, correlations between matrix elements require inserting appropriate weight factors for each pairing, set by the matrix of variances. Then, the calculation of a given moment can be performed as follows:
\begin{align}
    \langle \lambda^{2p}\rangle = \frac{1}{(I_d(\xi))^{p}}& \int_{\Lambda^{p+1}} \frac{d^{p+1}\mathbf{k}}
    {(2\pi)^{p+1}}\\& \times\sum_{\sigma \in NC([2p])}  \prod_{(k_i,k_j)\in\sigma} \hat G_d(k_i,k_j),
\end{align}
where $NC([2p])$ is the set of all non-crossing pairings of matrix entries. 

We illustrate this procedure by explicitly calculating the fourth and sixth moments. The fourth moment has a single $O(1)$ term, corresponding to diagram (4-ii) in Fig.~\ref{fig:pairings}, with a combinatorial factor of 2: 
\begin{equation}
    \langle\lambda^4\rangle = \frac{2}{I_d(\xi)^2} \int_{\Lambda^3} \frac{d^3 \mathbf{k}}{(2\pi)^3} \, \hat{G}_d(k_1, k_2) \, \hat{G}_d(k_1, k_3).
    \label{eqn:4momentintegral}
\end{equation}
The sixth moment can be calculated in a similar way by counting the contributions of diagrams (6-ii) and (6-iii) in Fig.~\ref{fig:pairings}: 
\begin{align}
    \langle\lambda^6\rangle =   \frac{1}{I_d(\xi)^3}\int_{\Lambda^4} \frac{d^4 \mathbf{k}}{(2\pi)^4} \,  3\hat{G}_d(k_1, k_2) \, \hat{G}_d(k_2, k_3) \, \hat{G}_d(k_3, k_4)  \nonumber\\ + 2 \hat{G}_d(k_1, k_2) \, \hat{G}_d(k_1, k_3) \,\hat{G}_d(k_1, k_4) \label{eqn:6momentintegral}.
\end{align}
The leading small $\xi$ behaviour of the relevant integrals can be calculated explicitly:
\begin{align}
\lim_{\xi\rightarrow0}  \langle\lambda^4\rangle =  2 + \frac{8\pi^4}{45}\xi^4; \label{eqn:leadinglambda4}\\
    \lim_{\xi\rightarrow0}  \langle\lambda^6\rangle =  5 + \frac{16\pi^4}{15}\xi^4.\label{eqn:leadinglambda6}
\end{align}
Equations~(IV.8) and (IV.9) describe only the leading departure from
the semicircle law in the limit $\xi\ll1$. For larger $\xi$, higher
orders contribute, and the full expressions in Eqs.~(IV.6) and
(IV.7) must be evaluated.
\begin{figure}[b]
  \centering
  \begin{subfigure}[b]{0.48\textwidth}
    \centering
    \includegraphics[width=\linewidth]{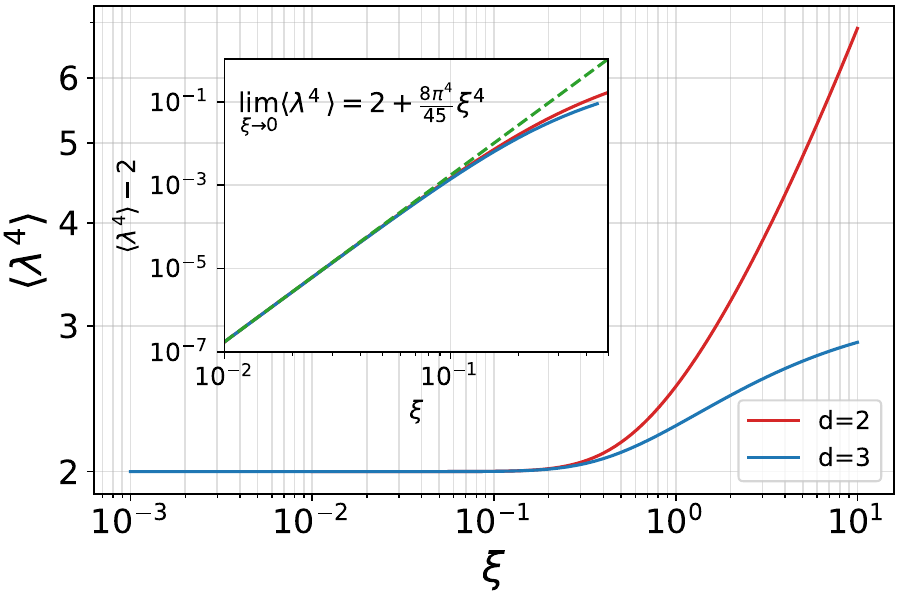}
  \end{subfigure}
  \hfill
  \begin{subfigure}[b]{0.48\textwidth}
    \centering
    \includegraphics[width=\linewidth]{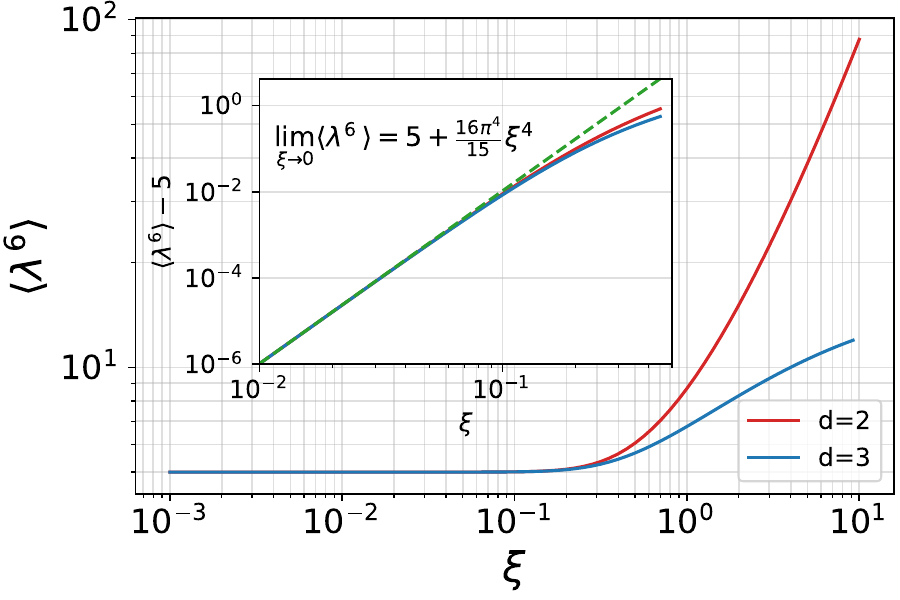}
  \end{subfigure}
  \caption{Fourth and sixth moments of the eigenvalue spectrum for $T>T_c$ as a function of $\xi$. The $d=2$ and $d=3$ curves show the full numerical evaluation of
eqs.~\eqref{eqn:4momentintegral} and \eqref{eqn:6momentintegral}. The dashed curves in the insets show the
leading small-$\xi$ results as indicated in eq.~\eqref{eqn:leadinglambda4}-\eqref{eqn:leadinglambda6}.}
  \label{fig:4n6moment}
\end{figure}
Numerically evaluating these expressions in the $T>T_c$ regime as functions of $\xi$ yields the results shown in Fig.~\ref{fig:4n6moment}. For $d=2$, the higher-order moments increase without bound as $\xi$ increases. As a result, regardless of the existence of a phase transition, the moments of the spectrum continue to increase with the correlation length. In contrast, for $d \geq 3$, the moments saturate to a constant value as $\xi \rightarrow \infty$. Evaluating the relevant integrals in the $\xi \rightarrow \infty$ limit, we obtain, for $d = 3$ and $d = 4$:
\begin{align}
    \lim_{\xi \rightarrow \infty} \langle \lambda^4\rangle \approx 3.17, \lim_{\xi \rightarrow \infty} \langle \lambda^6\rangle \approx 16.62 \text{ for } d = 3; \\
        \lim_{\xi \rightarrow \infty} \langle \lambda^4\rangle \approx 2.27, \lim_{\xi \rightarrow \infty} \langle \lambda^6\rangle \approx 6.63 \text{ for } d = 4.
\end{align}
These constants were obtained by numerical evaluation of the fully
normalized Brillouin-zone integrals in Eqs.~(IV.6) and (IV.7) in the
limit $\xi\rightarrow\infty$. The integration is performed using Mathematica's GlobalAdaptive rule, which ensures that the largest estimated error satisfies the precision goal. Convergence was verified by logarithmically increasing the correlation length value and the working precision. The results were stable well beyond the two decimal places reported. We have not found a useful closed-form
expression for these multidimensional integrals. As the dimension is further increased, the limiting value of the moments approaches the semicircle limit provided in eq.~\eqref{eqn:semicirclemoment}. Our method can be straightforwardly extended to higher-order moments using the procedure outlined above.

For $T=T_c$, the finite-size scaling of the moments is controlled by the scaling dimension $\Delta_\phi=(d-2+\eta)/2$ of the order-parameter field~\cite{Cardy_1996}. Power counting based on this long-distance power-law form shows that all contributions to a given moment, including both Wick pairings and higher connected correlators, contribute at the same leading order in system size. Using the normalization of the propagator derived in Section~\ref{sec:norm}, the covariance-controlled contribution has the scaling form
\begin{align}
    I_{2p}(L) \sim & L^{p} \int_{\Lambda^{p+1}} d^{p+1} \mathbf{k}\frac{1}{L^p|\mathbf{k}|^{(4-d-\eta)p}}, \nonumber \\
    \sim & \begin{cases}
    L^{(3-d-\eta)p-1}, & \eta < 3-d-1/p, \\
    \ln L, & \eta = 3-d-1/p, \\
    O(1), & \eta > 3-d-1/p.
    \end{cases} \label{eqn:powermoment}
\end{align}
The $2p$-th moment therefore grows algebraically when $\eta < 3-d-1/p$, logarithmically at the marginal value $\eta = 3-d-1/p$, and remains finite above it. For $d=2$, the threshold is $\eta_p=1-1/p$, while for $d=3$ it is $\eta_p=-1/p$. All fixed-order moments are therefore finite if $d+\eta\geq3$. Thus, as the dimension of the system increases, the signatures of long-range correlations and criticality in the eigenvalue spectrum become less distinct. For $d=2$, eq.~\eqref{eqn:powermoment} gives
\begin{equation}
    \langle \lambda^{2p}\rangle \sim L^{(1-\eta)p-1}, \qquad \eta<1-\frac{1}{p},
\end{equation}
with logarithmic growth at the marginal value $\eta=1-1/p$. In particular, for $p=2$,
\begin{equation}
    \langle \lambda^{4}\rangle \sim \begin{cases}
    L^{1-2\eta}, & \eta<1/2,\\
    \ln L, & \eta=1/2,\\
    O(1), & \eta>1/2.
    \end{cases} \label{eqn:fourthmomentscaling}
\end{equation}
For every finite matrix size $L$, all spectral moments are finite
because the matrix entries are bounded. At criticality, however, a
finite-$L$ moment may grow with system size and may therefore diverge
in the thermodynamic limit. This is the meaning of the scaling
relations in eqs.~\eqref{eqn:powermoment}--\eqref{eqn:fourthmomentscaling}.
Using the identification $\eta=2H$, the $d=2$ result becomes $\langle\lambda^{2p}\rangle\sim L^{p(1-2H)-1}$ for $H<(p-1)/(2p)$, with logarithmic growth at $H=(p-1)/(2p)$. This matches the finite-size moment hierarchy $H_p=(p-1)/(2p)$ found in Ref.~\cite{saberi2026longrangecorrelatedrandommatrices}. In particular, $\langle\lambda^4\rangle\sim L^{1-4H}$ for $H<1/4$, while at $H=1/4$ the fourth moment grows logarithmically.

\begin{figure*}[htbp]
  \centering
  \begin{subfigure}[b]{0.48\textwidth}
    \centering
    \includegraphics[width=\linewidth]{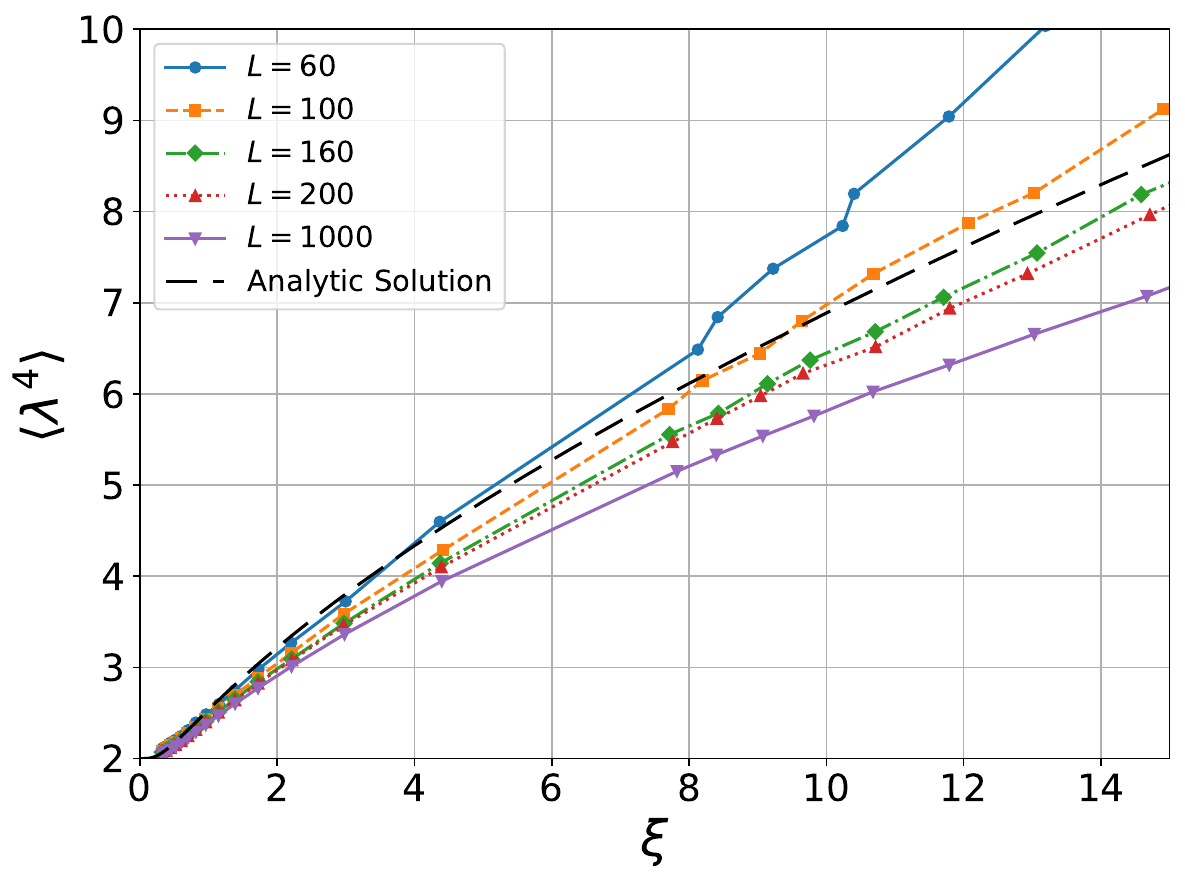}
  \end{subfigure}
  \begin{subfigure}[b]{0.48\textwidth}
    \centering
    \includegraphics[width=\linewidth]{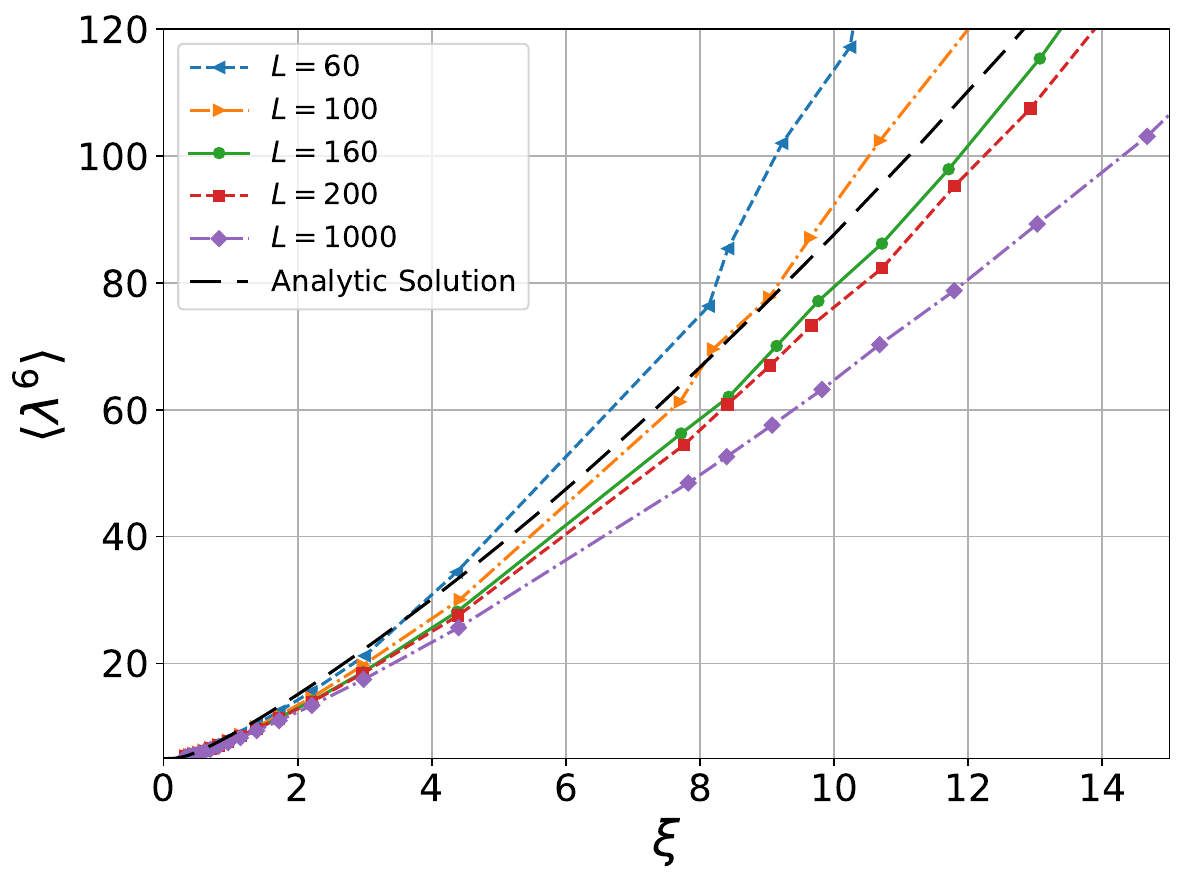}
  \end{subfigure}
  \begin{subfigure}[b]{0.48\textwidth}
    \centering
    \includegraphics[width=\linewidth]{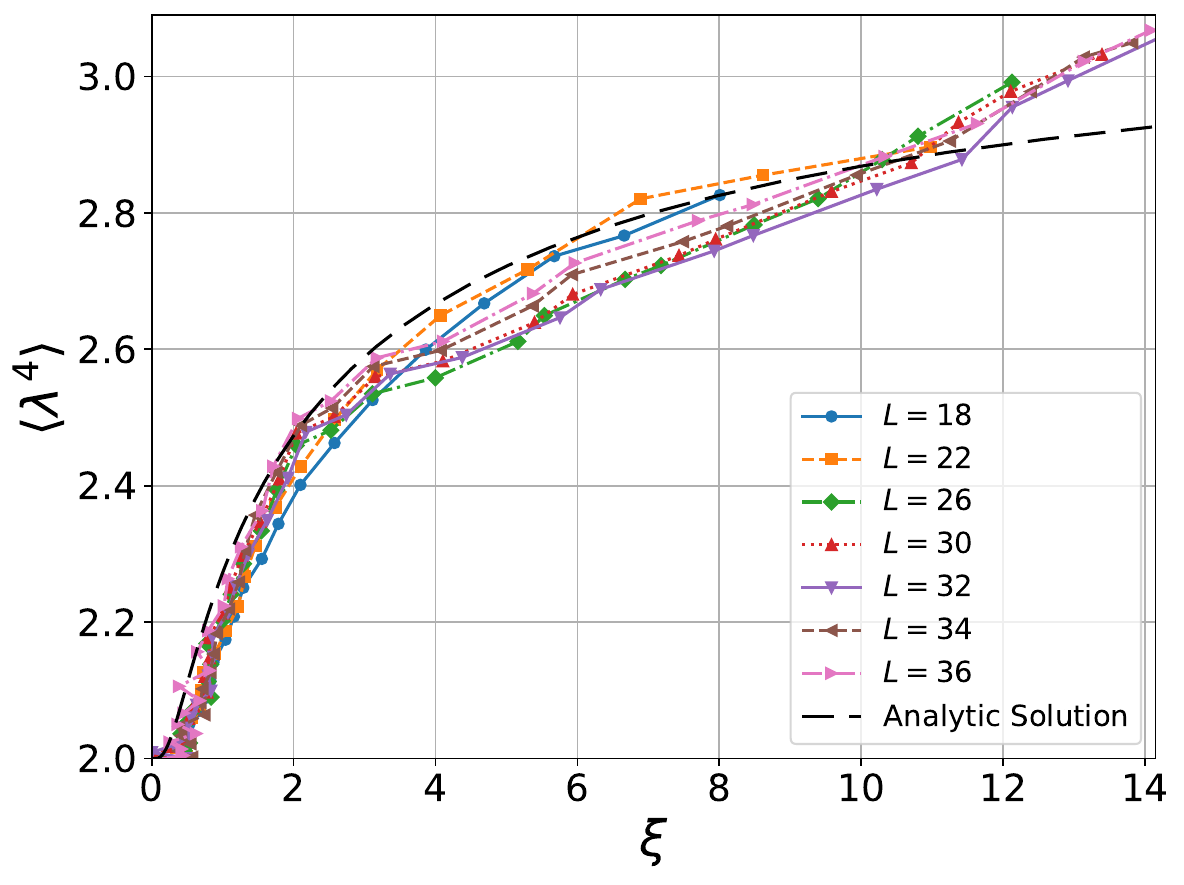}
  \end{subfigure}
  \begin{subfigure}[b]{0.48\textwidth}
    \centering
    \includegraphics[width=\linewidth]{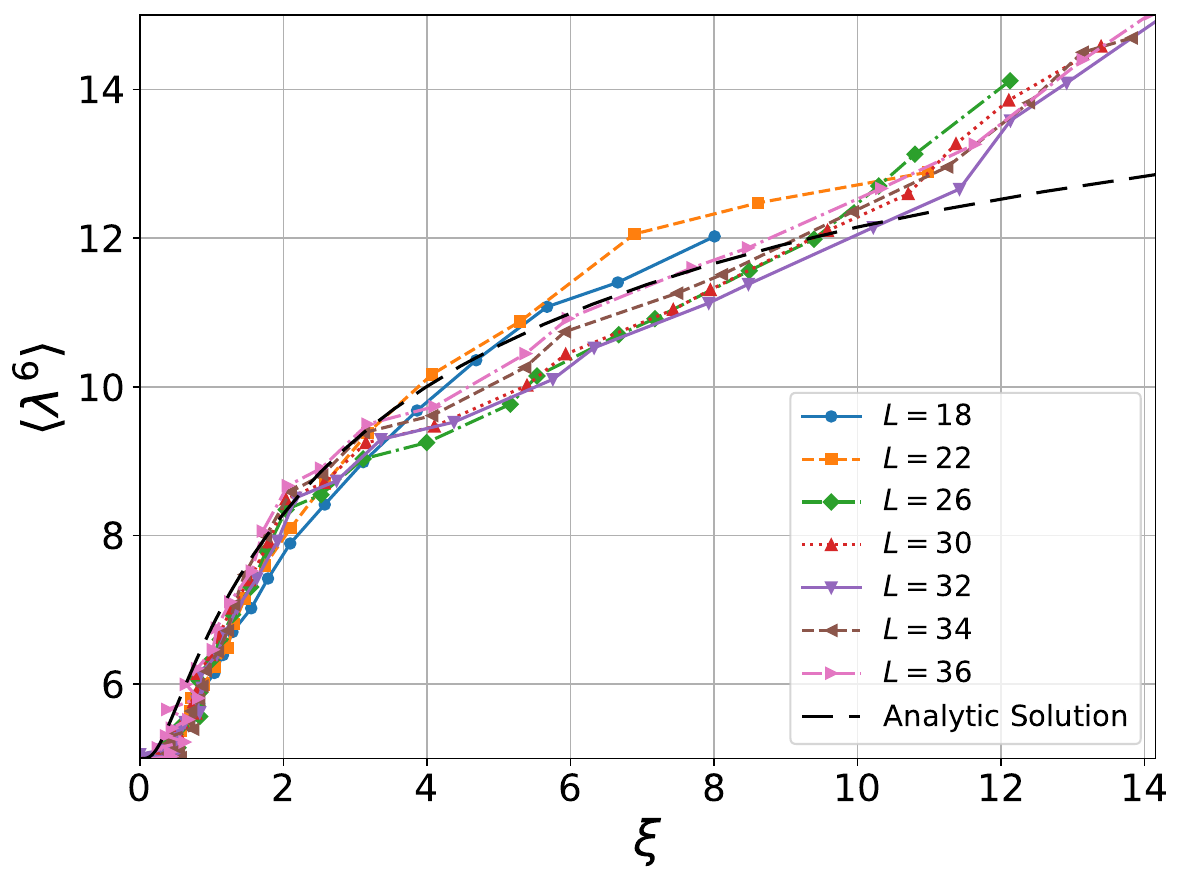}
  \end{subfigure}
  \caption{Fourth and sixth moments of the eigenvalue spectrum of Ising Model (top) and Gaussian EA spin glass (bottom) as a function of $\xi$, computed from Monte Carlo sampling. The black dashed line indicates the analytical solution shown in Fig.~\ref{fig:4n6moment}.}
  \label{fig:4n6momentMC}
\end{figure*}
Note that there is an emergent discontinuous transition in the values of the moments at criticality only if $\eta \neq 0$. For theories with $\eta = 0$, the Gaussian fixed point being a prominent example, the values of the moments at criticality coincide with those obtained by the $\xi \rightarrow \infty$ limit of the off-critical propagator. In this case, the behavior is fully determined by the two-point function, and higher-order cumulants do not contribute to the moment sum. For a Gaussian theory, we expect all moments to be finite for $d \geq 3$.

\begin{figure*}[htbp]
  \centering
  \begin{subfigure}[b]{0.48\linewidth}
    \centering
    \includegraphics[width=\linewidth]{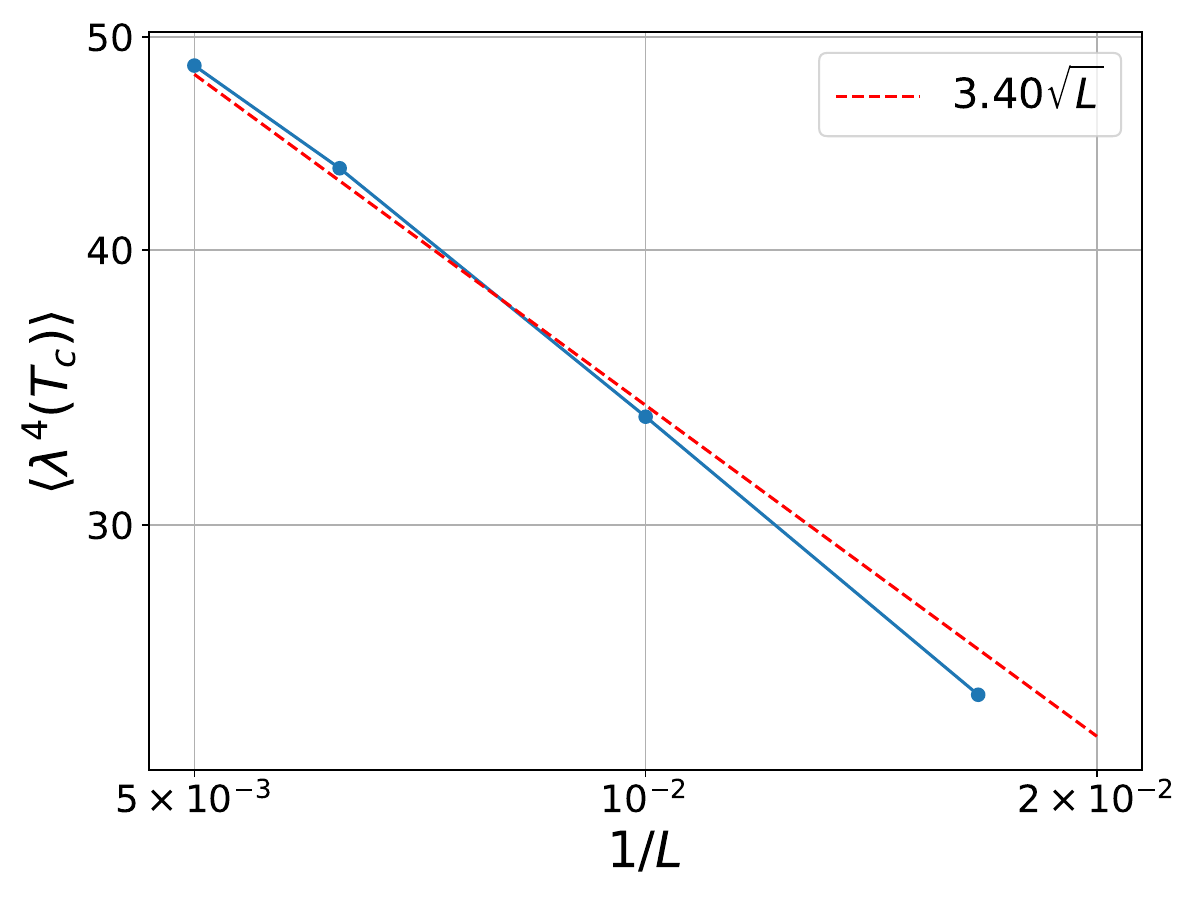}
  \end{subfigure}
  \hfill
  \begin{subfigure}[b]{0.48\linewidth}
    \centering
    \includegraphics[width=\linewidth]{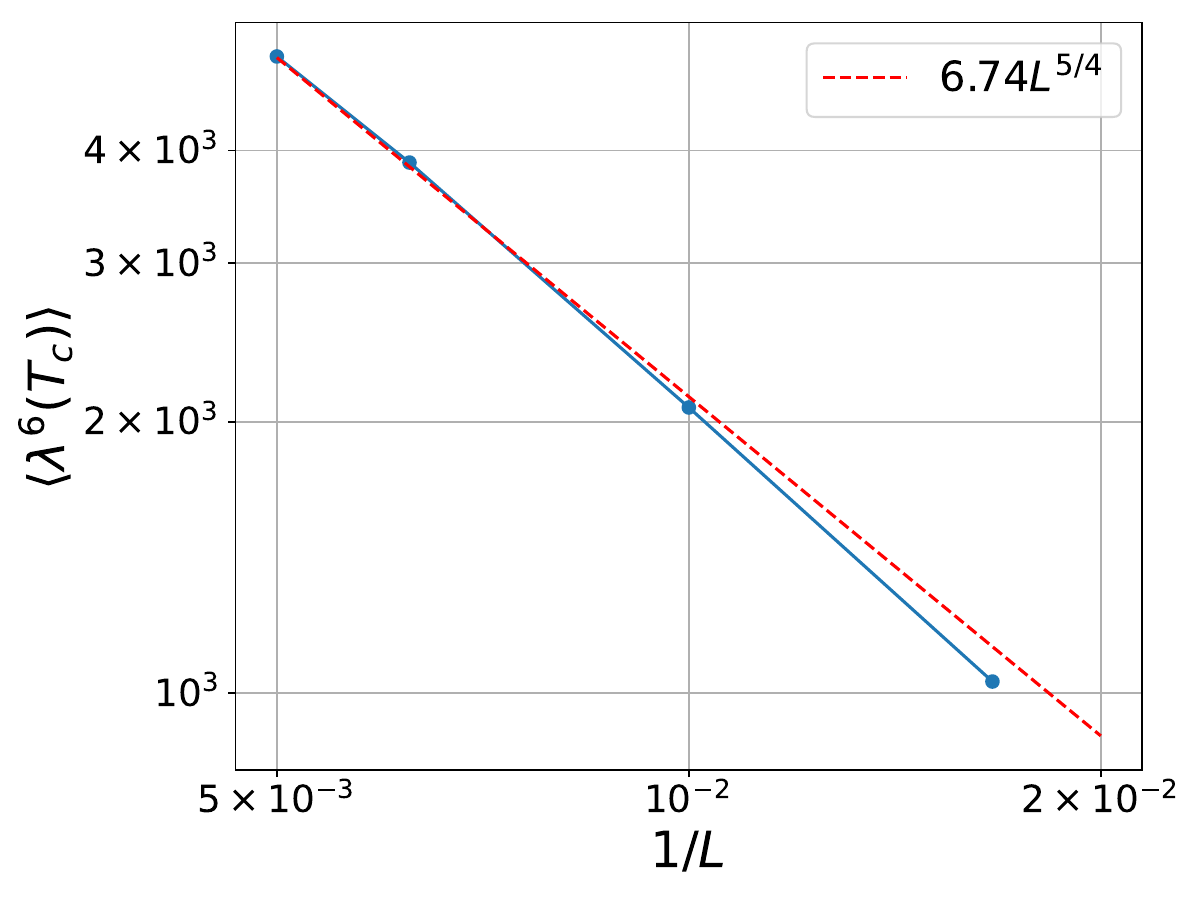}
  \end{subfigure}
  \caption{Scaling of the fourth and sixth moments of the eigenvalue spectrum at $T=T_c$ for the Ising model as a function of system size, computed from Monte Carlo sampling. The results show good agreement with the scaling predictions from a moment method analysis, and the results for the fourth moment agree with~\cite{Saberi2024}.}
  \label{fig:crit4n6moment}
\end{figure*}

In Fig.~\ref{fig:4n6momentMC}, we plot the behavior of the fourth and sixth moments of the Ising model and the 3D Edwards--Anderson spin glass as a function of their correlation length, and compare it with our theoretical results. For the Gaussian EA model, observables were averaged over 200 independent
disorder realizations. The correlation length was extracted by fitting
the disorder-averaged real-space decay of the overlap correlation
function. This procedure was found to be more stable than fitting the
momentum-space variance profile for the available system sizes. Our Monte Carlo results show good qualitative agreement with our analytical predictions. The analytical calculations can be made more accurate by including higher-order corrections to the Ornstein-Zernike form as calculated in~\cite{Mart_n_Mayor_2002}.
For the Ising model, $\eta=1/4$, we would expect the higher-order moments to diverge as $\langle \lambda^{2p}\rangle \sim L^{3p/4-1}$. In particular, we expect $\langle \lambda^{4}\rangle \sim L^{1/2}$, which explains the scaling behavior of the variance and kurtosis observed in~\cite{Saberi2024}. The scaling behavior of the fourth and sixth moments is shown in Fig.~\ref{fig:crit4n6moment}, which demonstrates good agreement with these predictions. 

For an EA spin glass with $\eta \sim -0.39$, our calculations imply a finite fourth moment and a slowly divergent sixth moment $\langle\lambda^{6} \rangle \sim L^{0.17}$. For the spin glass, we are not able to access system sizes large enough to stringently test the expected scaling behavior.

Given the method outlined in this section, a natural question to ask is whether the eigenvalue spectrum can be reconstructed given access to a finite sequence $\{\langle\lambda^k\rangle\}$ of its moments. This is known as the \textit{truncated moment problem}. Given a moment sequence, one can construct a family of polynomials $p_k(\lambda)$ which are orthogonal with respect to the inner product induced by the moments. Techniques based on orthogonal polynomials and Gaussian quadrature provide discrete approximations that match the moments up to a given order and weakly converge as the number of moments is increased if the underlying distribution is determinate~\cite{schmudgen2017moment}. Sufficient conditions for determinacy include support on a bounded interval or the Carleman condition $ \sum_{p=1}^\infty \langle \lambda^{2p} \rangle^{-\frac{1}{2p}} = +\infty.$~\cite{carleman}. 
\section{The Vector Dyson Equation of the Resolvent\label{sec:resolvent}}
\begin{figure*}
    \centering
    \begin{subfigure}{\linewidth}
        \centering
        \includegraphics[width=\linewidth]{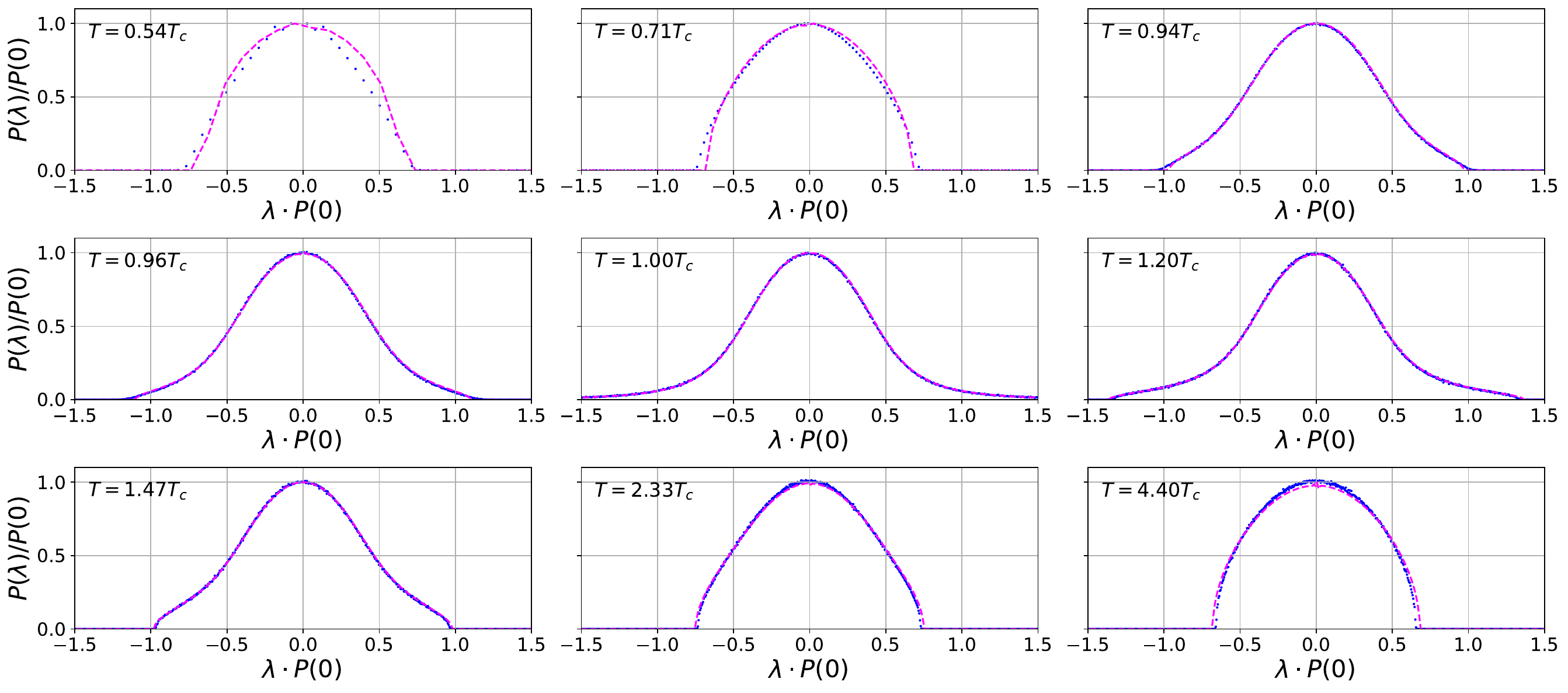}
            \captionsetup{justification=centering}
        \caption{}
        \label{fig:center_ising_compare}
    \end{subfigure}

    \vspace{0.5em}

    \begin{subfigure}{\linewidth}
        \centering
        \includegraphics[width=\linewidth]{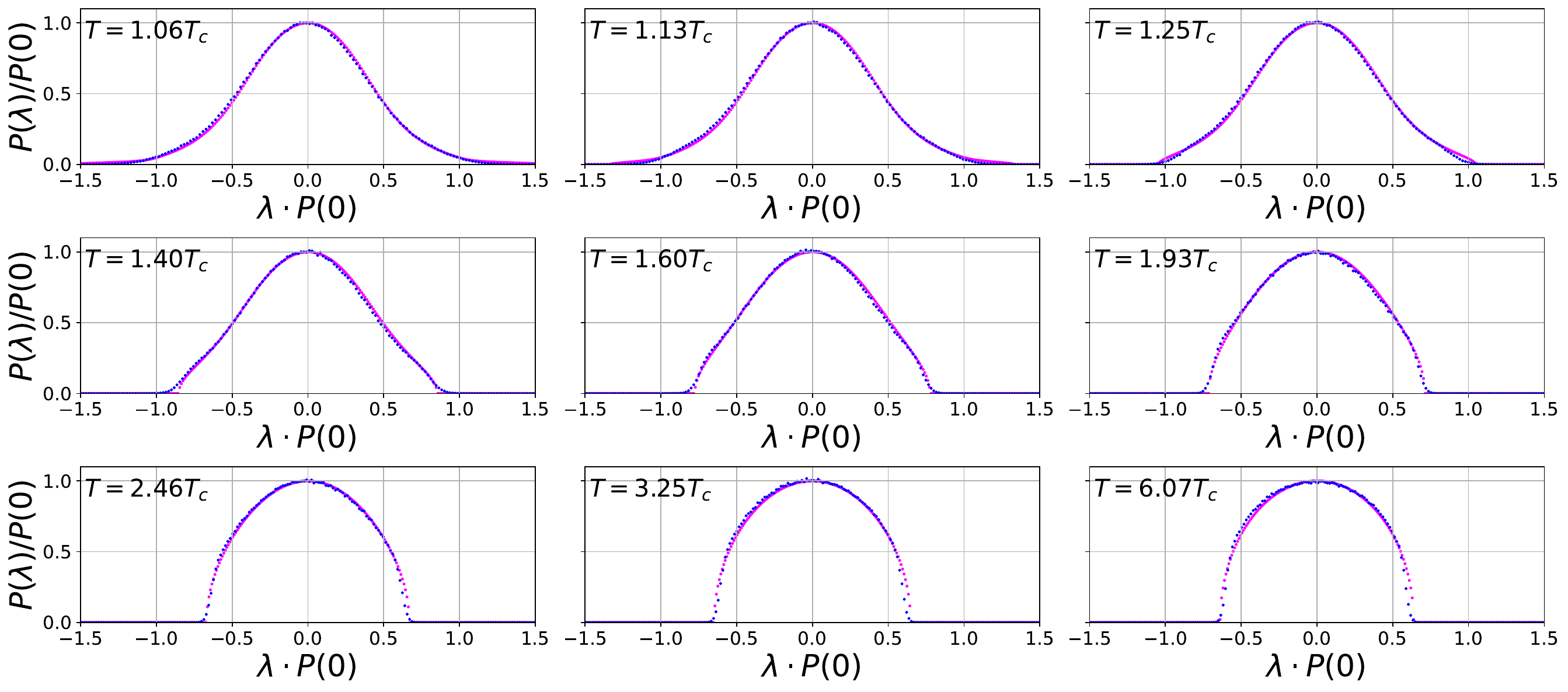}
            \captionsetup{justification=centering}
        \caption{}
        \label{fig:center_ea_compare}
    \end{subfigure}

    \caption{Comparison between Monte Carlo data (blue) and self-consistent solution from the Vector Dyson Equation (magenta). For visual comparison, we use
$x=\lambda P(0)$ and
$\widetilde P(x)=P(\lambda)/P(0)$. This gives
$\widetilde P(0)=1$ while preserving
$\int \widetilde P(x)\,dx=1$. The self-consistent solution is in good agreement with the bulk eigenvalue spectrum across all temperatures. (a) 2D Ising model of linear size $L=100$. The spectrum returns to the semicircle law as the temperature is decreased below $T = T_c$. (b) 3D Gaussian EA model of linear size $L=36$.}
    \label{fig:center_combined_compare}
\end{figure*}
Another technique for constructing the eigenvalue spectrum involves solving for the entire spectrum self-consistently without the need to explicitly calculate the moments. We will study the resolvent of a matrix $\mathcal{M}$, defined as 
\begin{equation}
    \mathcal{G}(z) = \frac{1}{\mathcal{M}-z}, z\in\mathbb{C}.
\end{equation}
By spectral decomposition,
\begin{equation}
    \mathcal{G}(z) = \sum_i \frac{\boldsymbol\nu_i\boldsymbol\nu_i^T}{\lambda_i-z},
\end{equation}
where $\mathcal{M}\boldsymbol\nu_i =\lambda_i\boldsymbol\nu_i$. It will also be helpful to define a vector of diagonal resolvent elements $\mathbf{m}_i=\mathcal G_{ii}$. Through a Stieltjes transform of the empirical eigenvalue distribution, we can connect the eigenvalue spectrum to the joint distribution of individual matrix elements. We start by defining the empirical measure of the eigenvalues:
\begin{equation}
P_L(d\lambda)
=
\frac{1}{L}\sum_{i=1}^L \delta(\lambda-\lambda_i)\,d\lambda .
\end{equation}
The Stieltjes transform of the empirical measure is given by
\begin{align}
   m(z)
   &=
   \int_{\mathbb{R}}\frac{P_L(d\lambda)}{\lambda-z},
   =
   \frac{1}{L}\sum_{i=1}^{L}\frac{1}{\lambda_i-z}
   \nonumber\\
   &=
   \frac{1}{L}\operatorname{Tr}\left(\frac{1}{\mathcal{M}-z}\right),
   \qquad z\in\mathbb{C}^{+}.
\end{align}
where $\mathbb{C}^{+}$ denotes the upper half plane. The Stieltjes transform can then be inverted to recover the eigenvalue density
\begin{equation}
    P_L(\lambda)
    =
    \frac{1}{\pi}
    \lim_{\epsilon\rightarrow0^{+}}
    \operatorname{Im}\,m(\lambda+i\epsilon).
    \label{eqn:inversestieltjes}
\end{equation}
For Wigner-type matrices, it was shown in~\cite{Ajanki_2016,ajanki2017universalitygeneralwignertypematrices,Ajanki_2019} that the deterministic limiting density of states is governed by a self-consistent equation determined by the variance matrix, known as the \textit{Vector Dyson Equation}:
\begin{equation}
    ((\mathcal{V}m(z))_i+z)m_i(z) + 1 = 0, i=1,...L .\label{eqn:vectordyson}
\end{equation}
Rigorous results establishing this behavior assume that the variance matrix satisfies~\cite{ajanki2017universalitygeneralwignertypematrices}
\begin{align}
    \langle \mathcal{M}_{ij} \rangle = 0, \frac{c}{L}\leq  \mathcal{V}_{ij} \leq \frac{C}{L} \forall (i,j) \in [1,L]^2, \label{eqn:vdcondition}
\end{align}
Under these assumptions, the self-consistent solution defines a compactly supported limiting density of states $P(\lambda)$ through the Stieltjes transform
\begin{equation}
\frac{1}{L}\sum_{i=1}^L m_i(z)
=
\int_{\mathbb{R}} \frac{P(\lambda)d\lambda}{\lambda - z},
\label{eqn:stieltjesP}
\end{equation}
with compact support on a finite interval whose scale is controlled by the variance profile:
\begin{equation}
    \Sigma = 2 \sqrt{||\mathcal{V}||_\infty}
    ,
\end{equation}
where $||\mathcal{V}||_\infty$ is the operator norm induced by the $\infty$-vector norm,
\begin{align}
||\mathcal{V}||_\infty
=
\max_{m}
\left(
\sum_{n=1}^L|\mathcal{V}_{mn}|
\right).
\label{eqn:infnorm}
\end{align}
In the rigorous Wigner-type setting this quantity controls the scale of the limiting support. In the critical regimes considered below, where the uniform $1/L$ bounds may fail, we use the same quantity as a diagnostic spectral scale rather than as a rigorous edge.

The self-consistent solution describes the eigenvalue density on macroscopic spectral scales, where $P(\lambda)=O(1)$. When $P(\lambda)\sim 1/L$, corresponding to microscopic spectral scales, local fluctuations play a dominant role in the eigenvalue statistics~\cite{ErdosYau2012}. Accordingly, the Vector Dyson equation is expected to accurately describe the bulk eigenvalues, while it does not generally control extreme or outlier eigenvalues~\cite{erdos2011bulkuniversalitygeneralizedwigner}.

The condition in eq.~\eqref{eqn:vdcondition} may fail at criticality, where the central mode scales as $\mathcal{V}_{00}\sim L^{3-d-\eta}$ (see eq.~\eqref{eqn:V00}). This breakdown of the uniform $1/L$ scaling reflects the emergence of a dominant mode in the variance profile, which can lead to the maximal eigenvalue separating from the bulk. In the critical regime, the Vector Dyson Equation is employed as a heuristic self-consistent approximation for the bulk spectral density rather than a rigorous limiting solution. 

This distinction is also consistent with the behavior observed in the long-range correlated ensemble of Ref.~\cite{saberi2026longrangecorrelatedrandommatrices}. There, decreasing $H$ strengthens the imposed long-range correlations and progressively enhances the infrared covariance. For $0<H<1/2$ the limiting density has algebraic tails; at $H=1/2$ its support remains unbounded even though every fixed moment is finite; and for $H>1/2$ the support is bounded, with the semicircle recovered as $H\to\infty$. For the specific covariance used in that work, the excess kurtosis crosses zero at $H_0=0.716556\ldots$, but this is not a Gaussian point because the sixth cumulant remains nonzero. These results separate the role of the infrared covariance, which controls the moment hierarchy and asymptotic tail, from that of the full covariance, which controls the detailed bulk density; they do not concern microscopic edge statistics.

In the present setting, the covariance law~\eqref{eqn:matvarcor} contains an additional term coupling Fourier modes of opposite sign. In the Gaussian translation-invariant case, it was shown in~\cite{Ajanki_2016} that this term does not affect the limiting global density of states. We will therefore use the Vector Dyson Equation~\eqref{eqn:vectordyson} to describe the global eigenvalue density of the ensembles considered here.

In addition to providing analytical tractability to the properties of the eigenvalue spectrum, the Vector Dyson Equation formalism provides an order-of-magnitude reduction in obtaining the eigenvalue spectrum in comparison to a direct eigenvalue sampling. The computational cost of eigenvalue decomposition on symmetric matrices scales as $O(L^\omega \ln^2(L/\epsilon))$ where $\omega \lesssim 2.371$ using state-of-the-art eigensolvers ~\cite{demmel2026minimizingarithmeticcommunicationcomplexity,sobczyk2025deterministiccomplexityanalysishermitian}. On the other hand, the complexity of Fourier transforming a 2D $L\times L$ matrix is $O(L^2\ln(L))$ via a discrete FFT~\cite{vanloan1992fft}. After computing the matrix of variances by sampling the Fourier-transformed ensemble, the Vector Dyson Equation needs to be solved once. Thus, this provides a significant computational advantage over direct sampling of the eigenvalues.

We apply this technique to the 2D Ising model and the 3D Edwards-Anderson spin glass by comparing spectra obtained from Monte Carlo sampling with those obtained from solving the Vector Dyson Equation using the variance matrix $\mathcal V$. The EA Monte Carlo spectra shown here are obtained from the simulation
data set used in Ref.~[13]. Reference~[13] studied a phenomenological
$q$-Gaussian description of these spectra. Here, the data is
used to test the variance-profile, moment, and Vector Dyson Equation
framework developed in the present work.

The variance matrix is computed by evaluating correlations in the Fourier-transformed ensemble, after which the Vector Dyson Equation is solved using an iterative relaxation scheme. We compare the self-consistent solution to the spectrum of all eigenvalues except the maximal eigenvalue, which separates from the bulk as $T\rightarrow T_c$ and is not captured by the self-consistent solution. The results are shown in Fig.~\ref{fig:center_combined_compare}.

The self-consistent solution appears to provide a good approximation to the bulk eigenvalue across all temperatures, indicating that the Vector Dyson Equation may still be used as a heuristic self-consistent approximation for systems in which the conditions listed in eq.~\eqref{eqn:vdcondition} fail.

\begin{figure}
    \centering
    \includegraphics[width=0.9\linewidth]{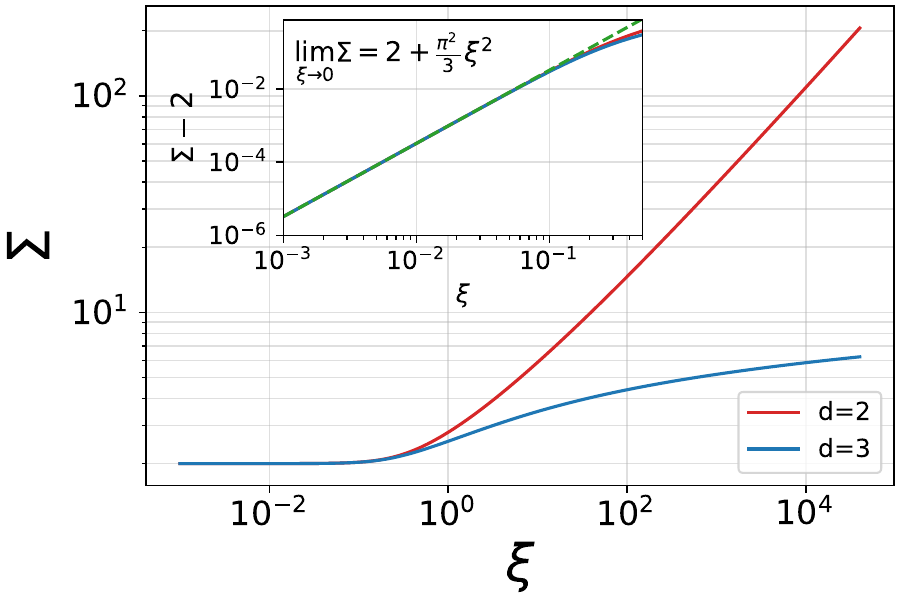}
    \caption{Spectral edge $\Sigma = 2\sqrt{||\mathcal{V}_{\infty}||}$ as a function of $\xi$ calculated for $d=2$ and $d=3$. The inset indicates the small $\xi$ behaviour.}
    \label{fig:spectraledge}
\end{figure}

We can use the formalism constructed in Section~\ref{sec:momentmethod} to calculate the value of the $\infty-$vector norm and the corresponding spectral edge for statistical physics random matrix ensembles. Having calculated the normalization, we evaluate the operator norm on the row in which $k_y=0$ where the maximal absolute row sum occurs. The contribution of the center element $\mathcal{V}_{00}=\chi/(\beta L^{d-1})$ scales as $\mathcal{V}_{00}\sim L^{3-d-\eta}f(tL^{1/\nu})$. Thus, in the $T>T_c$ regime, in the $L\rightarrow\infty$ limit, we will assume the contribution of this element is sub-leading. For $T>  T_c$, 
\begin{align}
 ||\mathcal{V}||_{\infty} & = \frac{1}{I_d(\xi)}  \int_{-\pi}^{\pi} \frac{1}{2\pi}dk \hat{G}_d((k,0),\xi), \label{eqn:sedgecal}
 \end{align}
which can be explicitly evaluated for $d=2$:
\begin{align}
  ||\mathcal{V}||_{\infty} = \frac{\xi\arctan{\pi \xi}}{\pi I_2(\xi) }  \text{ for } d=2.
\end{align}
In the $\xi\rightarrow 0$ limit for $T > T_c$, 
\begin{equation}
 \lim_{\xi \rightarrow0}||\mathcal{V}||_{\infty} =  1 + \frac{1}{3}\pi^2 \xi^2 + O(\xi^3).
\end{equation}
Thus, for $\xi=0$ we have recovered the semicircle value $\Sigma=2$.  The spectral edge $\Sigma(\xi)=2\sqrt {||\mathcal{V}(\xi)||_\infty}$ calculated from eq.~\eqref{eqn:sedgecal} is plotted for $d=2,3$ in Fig.~\ref{fig:spectraledge}. 
For $T<T_c$, the center mode has a macroscopic occupation, which will dominate the spectral norm:
\begin{align}
    ||\mathcal{V}||_{\infty} = L\langle q^2\rangle  \text{ for } T < T_c.
\end{align}

At $T=T_c$, a simple calculation with the correlator leads to a scaling law for the spectral edge
\begin{align}
    ||\mathcal{V}||_{\infty} & = \mathcal{V}_{00}+\frac{1 }{LI_d(\eta)}\int_{2\pi/L}^\pi \frac{L}{\pi}dk \frac{1}{k^{4-d-\eta}},
     \\ &  \sim 
        O(L^{3-d-\eta}).
\end{align}
\begin{figure*}
    \centering
    \begin{subfigure}[t]{0.48\linewidth}
        \centering
        \includegraphics[width=\linewidth]{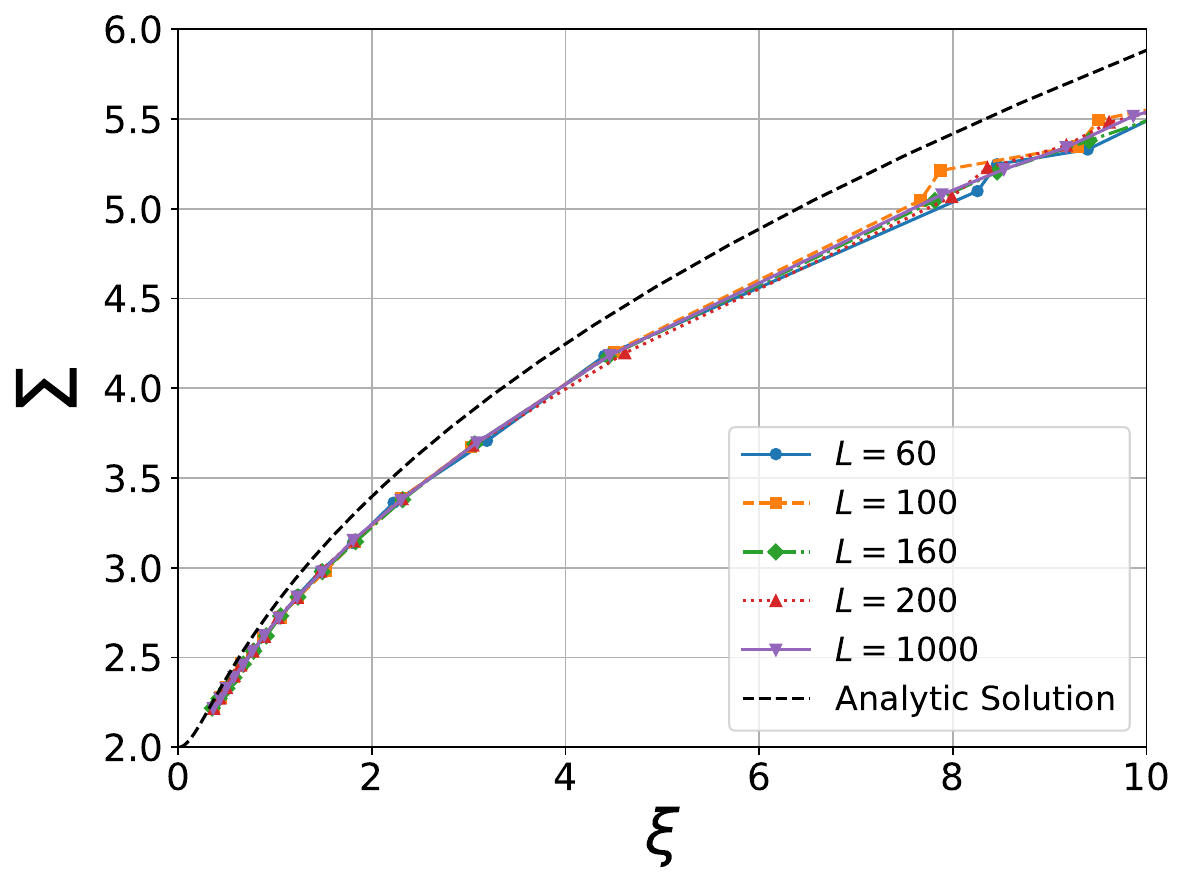}
    \end{subfigure}
    \hfill
    \begin{subfigure}[t]{0.48\linewidth}
        \centering
        \includegraphics[width=\linewidth]{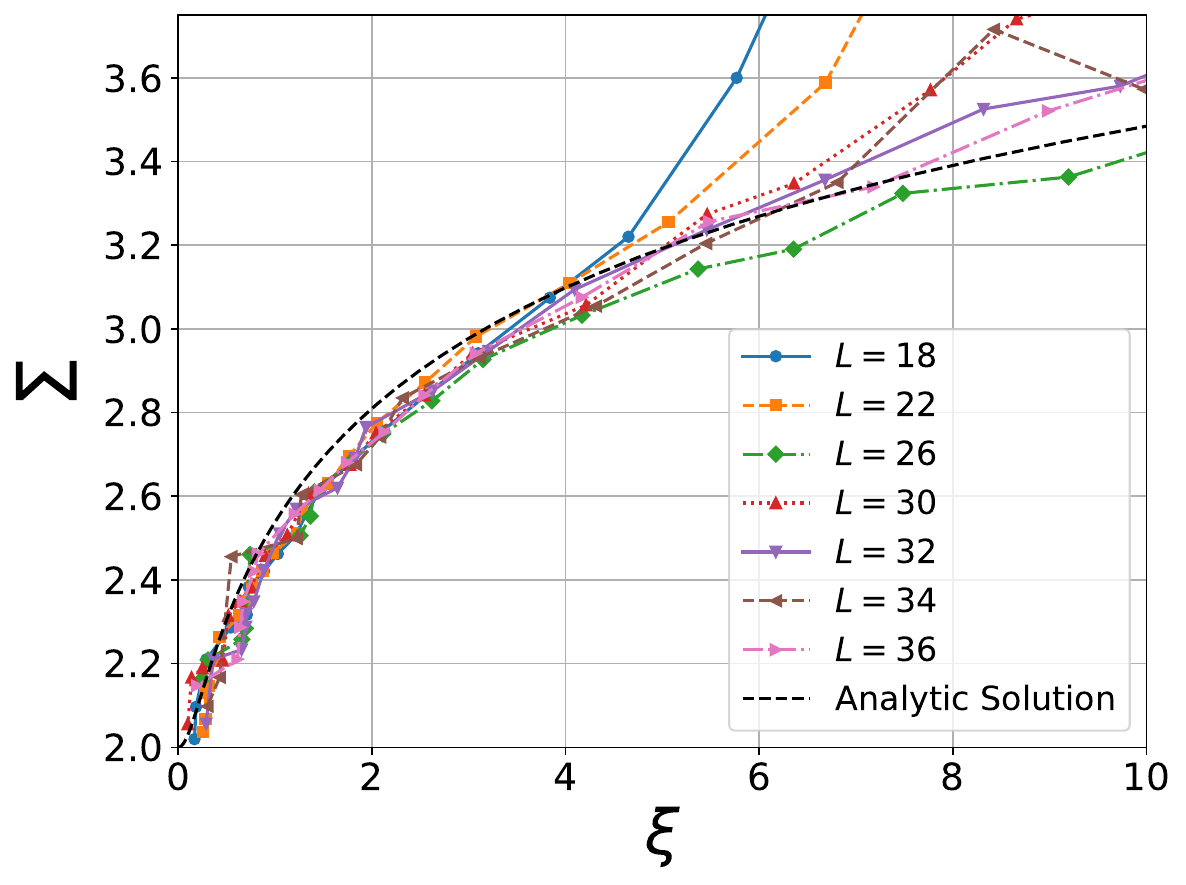}
    \end{subfigure}
    \caption{Spectral edge $\Sigma = 2\sqrt{||\mathcal{V}_{\infty}||}$ as a function of $\xi$, computed from Monte Carlo sampling of the Ising model (left) and the Gaussian EA spin glass (right). The analytic solution is plotted with the black dashed line.}
    \label{fig:SigmaMC}
\end{figure*}

Thus, a power-law decay in correlations can lead to a spectral scale that grows with system size if $d<3-\eta$. In such regimes, the variance-profile norm predicts a system-size-dependent broadening of the bulk spectral support. At a Gaussian fixed point $\eta=0$, this estimate gives a logarithmic growth for $d=3$ and a finite spectral scale for $d>3$. This analysis should be interpreted as a bulk-scale diagnostic: it is not expected to control microscopic edge fluctuations where the deterministic self-consistent solution is no longer the relevant description.

We assess the validity of our calculation for $||\mathcal{V}_{\infty}||$ by comparing our calculation for $\Sigma=2\sqrt{||\mathcal{V}_{\infty}||}$ with explicit results from Monte Carlo simulations for the Ising model and the EA spin glass in Fig.~\ref{fig:SigmaMC}. 
The analytical curve contains only the covariance-controlled Wick
contractions. The exact moments also contain connected four-point,
six-point, and higher-order cumulants of the underlying field. Their
sign is not fixed in a general interacting theory. In the range shown
here, the Wick approximation lies above the Monte Carlo data for the Ising model, implying a negative net contribution from the omitted terms. The EA spin glass, the curve appears to track the analytical prediction well until $\xi \sim 6$, after which the largest accessible system sizes indicate a small negative correction from higher-order cumulants.

\section{Extensions of the Framework}
\begin{figure}
    \centering
    \includegraphics[width=0.8\linewidth]{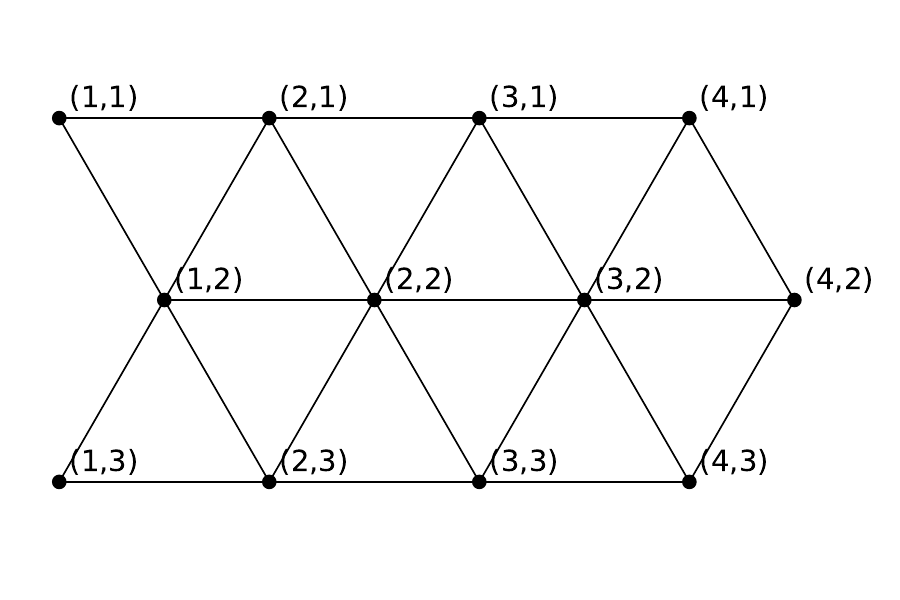}
    \caption{A statistical physics model on a triangular grid can be transformed into a random matrix ensemble by assigning matrix indices sequentially to vertices. An example index assignment is indicated for each vertex.}
    \label{fig:triangulargrid}
\end{figure}
In introducing the Short-Range Interaction-Correlated (\textbf{SIC}$(L,d,\xi)$) and Long-Range Interaction-Correlated (\textbf{LIC}$(L,d,\eta)$) random matrix ensembles, we have worked within a specific set of assumptions on the correlation structure of the underlying physical system. In the off-critical regime, we use a Gaussian/Wick approximation with a finite-correlation-length Ornstein--Zernike covariance profile, while at criticality we consider power-law correlations consistent with standard scaling behavior. These assumptions capture the generic behavior of a broad class of correlated systems while rendering the ensembles analytically tractable. The results presented here should therefore be viewed as a starting point for a more inclusive framework that can be systematically extended to less structured settings. In this section, we discuss natural extensions of our setup. 

The first level of generalization involves relaxing our assumptions in Section~\ref{sec:definition}. For instance, we have assumed that all odd moments of the matrix ensemble vanish due to invariance under sign flips of matrix elements. Introducing symmetry-breaking terms into the Hamiltonian, for instance, a magnetic field term for spin systems, will lead to non-symmetric eigenvalue spectra with nonzero odd moments. Applying the moment method in such cases would therefore require accounting for diagrams with an odd number of matrix insertions, as non-paired terms will no longer vanish. The standard Vector Dyson Equation framework, which relies on this symmetry, does not directly apply and would require modification. 

Another natural generalization involves relaxing the hypercubic lattice geometry assumption. For instance, one way to map a triangular system to a random matrix ensemble is to index into the vertices of each triangle as shown in Fig.~\ref{fig:triangulargrid}. The same method can be used to analyze a hexagonal system through a star-triangle identity~\cite{AuYangPerk1989}. While this preserves translational invariance and allows a Fourier-space formulation similar to Section~\ref{sec:definition}, the mapping does not preserve the original lattice topology. As a result, the correspondence between the correlation structure of the matrix ensemble and that of the underlying physical system becomes less direct. A natural direction for future work is to investigate whether the spectral observables of the resulting random matrix can capture critical behavior on non-rectangular geometries.

Our construction can be further extended to rectangular geometries by replacing eigenvalue decomposition with singular value decomposition. In this setting, the moment method can be adapted to singular values with modified combinatorial counting rules~\cite{bai2010spectral}. Following a rectangular Fourier transform of a correlated ensemble, an $m\times n$ matrix of variances determines the moments of the singular value distribution. A self-consistent solution can also be derived for the singular value distribution through a modification of the Vector Dyson Equation~\cite{fleermann2022proofmethodsrandommatrix}.

Our analysis here has been further restricted to thermal classical statistical ensembles, but the construction can be extended to quantum systems. Identifying suitable order parameters for quantum phase transitions remains a longstanding challenge, particularly in systems where conventional symmetry-breaking descriptions fail~\cite{Shi_2015,mariella2025orderparameterdiscoveryquantum}. Spectral properties of many-body operators, such as level statistics and the density of states, are well-established probes for distinguishing quantum phases~\cite{shklovskii,Evers_2008}. This suggests that random matrix ensembles derived from quantum systems may capture aspects of quantum critical behavior. A natural extension of this work, therefore, involves examining the effect of quantum fluctuations on the spectral observables of random matrix ensembles generated from quantum lattice systems. 

\section{Outlook}
We have presented a technique for mapping statistical physics systems to random matrices by directly translating the internal degrees of freedom to matrix elements. We have constructed two main techniques to characterize the eigenvalue spectrum of the resulting random matrix ensembles: the moment method, which calculates the moments of the eigenvalue spectrum, and the resolvent method, which provides an approximate self-consistent solution to the bulk eigenvalue spectrum. Together, these techniques provide a flexible analytical framework for estimating the eigenvalue spectra of statistical-physics random matrices from their underlying correlation structure. The analytical techniques we have presented can be applied to ordered and disordered models. When applied to systems in $d>2$ dimensions, our framework provides a natural way to reduce the model to an effective $d=2$ description while retaining salient features. 

Our findings demonstrate that our setup can be used to obtain a set of observables that naturally capture prominent features of a statistical physics system. Crucially, the eigenvalue spectrum is sensitive to short-range correlations, thus providing a natural probe to distinguish internal structure within the $T> T_c$ phase. Furthermore, the critical eigenvalue spectra are determined by the classification of the phase transition, thus establishing spectral observables as new probes of critical behavior. The long-range correlated ensemble of Ref.~\cite{saberi2026longrangecorrelatedrandommatrices} provides a limiting case in which the correlation exponent can be tuned directly. It shows that the long-distance exponent determines the hierarchy of moment thresholds and asymptotic spectral tails, whereas the full covariance determines the central density and finite shape measures. The present framework embeds this covariance-to-spectrum mechanism into physical lattice models: the scaling of the moments, shape of the spectral density, and the possible separation of low-momentum modes are now controlled by quantities naturally appearing in thermal contexts, such as $\xi$, $\eta$, $\chi$, and the order parameter. This suggests a broader classification program in which universality classes of statistical mechanics are associated not only with real-space critical exponents, but also with corresponding random matrix spectral universality classes.

Our approach is particularly promising when applied to statistical physics systems lacking an established analytical description of their phase transitions. For instance, our findings for the EA spin glass can provide additional insight into the contention between the RSB and Droplet viewpoints. In this regard, our method is a computationally efficient alternative to the multi-replica correlator methods typically used in Monte Carlo studies of spin glasses~\cite{Baity_Jesi_2013}. 

Another promising application of our framework lies at the intersection of learning theory and statistical physics. Learning theory gains significant insight from both statistical physics~\cite{star2024understandingmachinelearningparadigms} and random matrix theory~\cite{dandi2024randommatrixtheoryperspective}. Thus, statistical physics random matrices can inspire newer methodologies in learning theory. Conversely, our construction can be applied to efficiently train models for learning phase transitions. For instance, in~\cite{Wang_2016}, the singular values of a conserved-order-parameter Ising model were used to train an unsupervised neural network to recognize its phase transition. Our method offers a generalized, systematic way to quantify the behavior of the spectral components used in training these models. 

Although our focus has been on the global eigenvalue spectrum in this paper, eigenvalue statistics can be probed on smaller scales. At the microscopic scale corresponding to level spacings on the order $L^{-1}$, Wigner-type matrices conform to universal Wigner-Dyson statistics~\cite{ErdosYau2012,Ajanki_2018}. Our previous results on the eigenvalue spacing statistics on Edwards-Anderson random matrices are consistent with universal statistics, indicating that microscopic eigenvalue statistics are largely insensitive to correlations~\cite{sdbx-wx5t}. In contrast, mesoscopic statistics associated with widths $1\gg\delta P(\lambda) \gg L^{-1}$ are sensitive to the resolvent and may therefore be useful probes of correlation structure~\cite{erdos2019matrixdysonequationapplications}. Alternatively, observables defined through the eigenstates, such as the inverse participation ratio (IPR), can become informative in non-mean-field regimes, where localization or multifractal behavior may arise~\cite{Mirlin2000}.

\textit{Acknowledgments}---This work was supported by the Deutsche Forschungsgemeinschaft (DFG, German Research Foundation) under Project No.~557852701 (A.A.S.), and in part by Research Unit FOR~5522 (Project ID 499180199) and the Cluster of Excellence ct.qmat (EXC~2147, Project No.~390858490) (R.M.). Additional support was provided by the Advanced Study Group \emph{``Strongly Correlated Extreme Fluctuations''} at the Max Planck Institute for the Physics of Complex Systems, Dresden (2024/25)~\cite{pks_asg2024}.

\bibliography{references}

@misc{pks_asg2024,
  howpublished = {\url{https://www.pks.mpg.de/asg2024}}
}

@book{mehta1991random,
  title     = {Random Matrices},
  author    = {Mehta, Madan Lal},
  publisher = {Academic Press},
  address   = {New York},
  edition   = {2},
  year      = {1991}
}

@article{Baity_Jesi_2013,
   title={Critical parameters of the three-dimensional Ising spin glass},
   volume={88},
   ISSN={1550-235X},
   url={http://dx.doi.org/10.1103/PhysRevB.88.224416},
   DOI={10.1103/physrevb.88.224416},
   number={22},
   journal={Physical Review B},
   publisher={American Physical Society (APS)},
   author={Baity-Jesi, M. and Baños, R. A. and Cruz, A. and Fernandez, L. A. and Gil-Narvion, J. M. and Gordillo-Guerrero, A. and Iñiguez, D. and Maiorano, A. and Mantovani, F. and Marinari, E. and Martin-Mayor, V. and Monforte-Garcia, J. and Sudupe, A. Muñoz and Navarro, D. and Parisi, G. and Perez-Gaviro, S. and Pivanti, M. and Ricci-Tersenghi, F. and Ruiz-Lorenzo, J. J. and Schifano, S. F. and Seoane, B. and Tarancon, A. and Tripiccione, R. and Yllanes, D.},
   year={2013},
   month=dec }

@inbook{Altieri_2024,
   title={An introduction to the theory of spin glasses},
   ISBN={9780323914086},
   url={http://dx.doi.org/10.1016/B978-0-323-90800-9.00249-3},
   DOI={10.1016/b978-0-323-90800-9.00249-3},
   booktitle={Encyclopedia of Condensed Matter Physics},
   publisher={Elsevier},
   author={Altieri, Ada and Baity-Jesi, Marco},
   year={2024},
   pages={361–370} }

@book{Anderson_Guionnet_Zeitouni_2009, place={Cambridge}, series={Cambridge Studies in Advanced Mathematics}, title={An Introduction to Random Matrices}, publisher={Cambridge University Press}, author={Anderson, Greg W. and Guionnet, Alice and Zeitouni, Ofer}, year={2009}, collection={Cambridge Studies in Advanced Mathematics}}

@article{erdos2019matrixdysonequationapplications,
  title   = {The Matrix Dyson Equation and Its Applications for Random Matrices},
  author  = {Erd{\H{o}}s, L{\'a}szl{\'o}},
  journal = {arXiv preprint arXiv:1903.10060},
  year    = {2019},
  url     = {https://arxiv.org/abs/1903.10060}
}

@article{Ajanki_2016,
   title={Local Spectral Statistics of Gaussian Matrices with Correlated Entries},
   volume={163},
   ISSN={1572-9613},
   url={http://dx.doi.org/10.1007/s10955-016-1479-y},
   DOI={10.1007/s10955-016-1479-y},
   number={2},
   journal={Journal of Statistical Physics},
   publisher={Springer Science and Business Media LLC},
   author={Ajanki, Oskari H. and Erdős, László and Krüger, Torben},
   year={2016},
   month=feb, pages={280–302} }

@article{Ajanki_2019,
   title={Quadratic Vector Equations On Complex Upper Half-Plane},
   volume={261},
   ISSN={0065-9266},
   url={http://dx.doi.org/10.1090/memo/1261},
   DOI={10.1090/memo/1261},
   number={1261},
   journal={Memoirs of the American Mathematical Society},
   publisher={American Mathematical Society (AMS)},
   author={Ajanki, Oskari and Erdős, László and Krüger, Torben},
   year={2019},
   month=Nov }

@article{erdos2011bulkuniversalitygeneralizedwigner,
	author = {Erd{\H o}s, L{\'a}szl{\'o} and Yau, Horng-Tzer and Yin, Jun},
	journal = {Probability Theory and Related Fields},
	number = {1},
	pages = {341--407},
	title = {Bulk universality for generalized Wigner matrices},
	volume = {154},
	year = {2012},
    doi = { https://doi.org/10.1007/s00440-011-0390-3},
    
}

@article{ajanki2017universalitygeneralwignertypematrices,
	author = {Ajanki, Oskari H. and Erd{\H o}s, L{\'a}szl{\'o} and Kr{\"u}ger, Torben},
	journal = {Probability Theory and Related Fields},
	number = {3},
	pages = {667--727},
	title = {Universality for general Wigner-type matrices},
	volume = {169},
	year = {2017},
   doi = {https://doi.org/10.1007/s00440-016-0740-2}
    }

@book{goldenfeld,
	author = {Goldenfeld, N. },
	publisher = {Basic Books},
	title = {Lectures On Phase Transitions And The Renormalization Group},
	year = {1992}}

@article{Wang_2016,
   title={Discovering phase transitions with unsupervised learning},
   volume={94},
   ISSN={2469-9969},
   url={http://dx.doi.org/10.1103/PhysRevB.94.195105},
   DOI={10.1103/physrevb.94.195105},
   number={19},
   journal={Physical Review B},
   publisher={American Physical Society (APS)},
   author={Wang, Lei},
   year={2016},
   month=nov }

@article{Saberi2024,
   title={Interaction-correlated random matrices},
   volume={110},
   ISSN={2469-9969},
   url={http://dx.doi.org/10.1103/PhysRevB.110.L180102},
   DOI={10.1103/physrevb.110.l180102},
   number={18},
   journal={Physical Review B},
   publisher={American Physical Society (APS)},
   author={Saberi, Abbas Ali and Saber, Sina and Moessner, Roderich},
   year={2024},
   month=Nov 
   }

@article{Mart_n_Mayor_2002,
   title={Critical structure factor in Ising systems},
   volume={66},
   ISSN={1095-3787},
   url={http://dx.doi.org/10.1103/PhysRevE.66.026112},
   DOI={10.1103/physreve.66.026112},
   number={2},
   journal={Physical Review E},
   publisher={American Physical Society (APS)},
   author={Martín-Mayor, Victor and Pelissetto, Andrea and Vicari, Ettore},
   year={2002},
   month=aug }

@ARTICLE{OrnsteinZernike1914,
        title = "{Accidental deviations of density and opalescence at the critical point of a single substance}",
      journal = {Koninklijke Nederlandse Akademie van Wetenschappen Proceedings Series B Physical Sciences},
      author = {L.S. Ornstein and F. Zernike},
         year = 1914,
        month = jan,
       volume = {17},
        pages = {793-806},
       url = {https://dwc.knaw.nl/DL/publications/PU00012727.pdf} }

@book{schmudgen2017moment,
  title = {The Moment Problem},
  author = {Schmüdgen, Konrad},
  year = {2017},
  edition = {1st},
  publisher = {Springer International Publishing},
  address = {Cham},
  doi = {https://doi.org/10.1007/978-3-319-64546-9}}

@article{fleermann2022proofmethodsrandommatrix,
	author = {Michael Fleermann and Werner Kirsch},
	journal = {Probability Surveys},
	number = {none},
	pages = {291 -- 381},
	title = {{Proof methods in random matrix theory}},
	volume = {20},
	year = {2023},
    doi = {https://doi.org/10.1214/23-PS16}
    }

@book{bai2010spectral,
  author    = {Z. D. Bai and J. W. Silverstein},
  title     = {Spectral Analysis of Large Dimensional Random Matrices},
  series    = {Springer Series in Statistics},
  edition   = {2},
  year      = {2010},
  publisher = {Springer},
  doi       = {10.1007/978-1-4419-0661-8}
}

@incollection{AuYangPerk1989,
  author    = {Helen Au-Yang and Jacques H. H. Perk},
  title     = {Onsager’s Star-Triangle Equation: Master Key to Integrability},
  booktitle = {Advanced Studies in Pure Mathematics},
  volume    = {19},
  pages     = {57--94},
  year      = {1989},
  publisher = {Academic Press},
  doi       = {10.2969/aspm/01910057}
}

@article{star2024understandingmachinelearningparadigms,
  title   = {Understanding Machine Learning Paradigms through the Lens of Statistical Thermodynamics: A Tutorial},
  author  = {{Star (Xinxin) Liu}},
  journal = {arXiv preprint arXiv:2411.15945},
  year    = {2024},
  url     = {https://arxiv.org/abs/2411.15945}
}

@article{dandi2024randommatrixtheoryperspective,
  title   = {A Random Matrix Theory Perspective on the Spectrum of Learned Features and Asymptotic Generalization Capabilities},
  author  = {Dandi, Yatin and Pesce, Luca and Cui, Hugo and Krzakala, Florent and Lu, Yue M. and Loureiro, Bruno},
  journal = {arXiv preprint arXiv:2410.18938},
  year    = {2024},
  url     = {https://arxiv.org/abs/2410.18938}
}

@article{carleman,
  title={{\"U}ber die Abelsche Integralgleichung mit konstanten Integrationsgrenzen},
  author={Torsten Carleman},
  journal={Mathematische Zeitschrift},
  year={1922},
  volume={15},
  pages={111-120},
  url={https://api.semanticscholar.org/CorpusID:120060711}
}

@misc{demmel2026minimizingarithmeticcommunicationcomplexity,
      title={Minimizing the Arithmetic and Communication Complexity of Jacobi's Method for Eigenvalues and Singular Values: Part One -- Serial Algorithms}, 
      author={James Demmel and Hengrui Luo and Ryan Schneider and Yifu Wang},
      year={2026},
      eprint={2506.03466},
      archivePrefix={arXiv},
      primaryClass={math.NA},
      url={https://arxiv.org/abs/2506.03466}, 
}

@misc{sobczyk2025deterministiccomplexityanalysishermitian,
  title         = {Deterministic Complexity Analysis of Hermitian Eigenproblems},
  author        = {Sobczyk, Aleksandros},
  year          = {2025},
  eprint        = {2410.21550},
  archivePrefix = {arXiv},
  primaryClass  = {cs.DS},
  url           = {https://arxiv.org/abs/2410.21550}
}

@book{vanloan1992fft,
  author    = {Van Loan, Charles F.},
  title     = {Computational Frameworks for the Fast Fourier Transform},
  publisher = {Society for Industrial and Applied Mathematics},
  year      = {1992},
  series    = {Frontiers in Applied Mathematics},
  volume    = {10},
  doi       = {10.1137/1.9781611970999},
  isbn      = {978-0-89871-285-8}
}

@article{sdbx-wx5t,
  title={q-Gaussian Crossover in Overlap Spectra toward 3D Edwards-Anderson Criticality},
  author={{\"O}nder, Yaprak and Saberi, Abbas Ali and Moessner, Roderich},
  journal={Physical Review Letters},
  volume={136},
  number={8},
  pages={087103},
  year={2026},
  publisher={APS},
    doi = {10.1103/sdbx-wx5t},
  url = {https://link.aps.org/doi/10.1103/sdbx-wx5t}
}

@misc{anderson2007lawlargenumbersfiniterange,
      title={A law of large numbers for finite-range dependent random matrices}, 
      author={Greg Anderson and Ofer Zeitouni},
      year={2007},
      eprint={math/0609364},
      archivePrefix={arXiv},
      primaryClass={math.PR},
      url={https://arxiv.org/abs/math/0609364}, 
}

@article{Pastur1996Eigenvalue,
  author  = {L. Pastur},
  title   = {Eigenvalue distribution of random matrices: some recent results},
  journal = {Annales de L’I.H.P., Physique th\'eorique},
  volume  = {64},
  number  = {3},
  pages   = {325--337},
  year    = {1996},
  url = {https://www.numdam.org/item/AIHPA_1996__64_3_325_0.pdf}
}

@book{kardar2007statistical,
  title={Statistical Physics of Fields},
  author={Kardar, Mehran},
  year={2007},
  publisher={Cambridge University Press}
}

@article{WilsonKogut1974,
  author  = {Kenneth G. Wilson and J. Kogut},
  title   = {The renormalization group and the $\epsilon$ expansion},
  journal = {Physics Reports},
  volume  = {12},
  number  = {2},
  pages   = {75--199},
  year    = {1974},
  doi     = {10.1016/0370-1573(74)90023-4}
}

@article{Fisher1998,
  author  = {Michael E. Fisher},
  title   = {Renormalization group theory: Its basis and formulation in statistical physics},
  journal = {Reviews of Modern Physics},
  volume  = {70},
  number  = {2},
  pages   = {653--681},
  year    = {1998},
  doi     = {10.1103/RevModPhys.70.653}
}

@article{HohenbergHalperin1977,
  author  = {P. C. Hohenberg and B. I. Halperin},
  title   = {Theory of dynamic critical phenomena},
  journal = {Reviews of Modern Physics},
  volume  = {49},
  number  = {3},
  pages   = {435--479},
  year    = {1977},
  doi     = {10.1103/RevModPhys.49.435}
}

@book{LandauBinder2021,
  author    = {David P. Landau and Kurt Binder},
  title     = {A Guide to Monte Carlo Simulations in Statistical Physics},
  edition   = {5},
  publisher = {Cambridge University Press},
  address   = {Cambridge},
  year      = {2021},
  doi       = {10.1017/9781108780346}
}

@article{BerthierReichman2023,
  author  = {Ludovic Berthier and David R. Reichman},
  title   = {Modern computational studies of the glass transition},
  journal = {Nature Reviews Physics},
  volume  = {5},
  pages   = {102--116},
  year    = {2023},
  doi     = {10.1038/s42254-022-00548-x}
}

@article{CarrasquillaMelko2017,
  author  = {Juan Carrasquilla and Roger G. Melko},
  title   = {Machine learning phases of matter},
  journal = {Nature Physics},
  volume  = {13},
  pages   = {431--434},
  year    = {2017},
  doi     = {10.1038/nphys4035}
}

@article{ErdosYau2012,
  author  = {L{\'a}szl{\'o} Erd{\H{o}}s and Horng-Tzer Yau},
  title   = {Universality of local spectral statistics of random matrices},
  journal = {Bulletin of the American Mathematical Society},
  volume  = {49},
  number  = {3},
  pages   = {377--414},
  year    = {2012},
  doi     = {10.1090/S0273-0979-2012-01372-1}
}

@article{ZhangGopalakrishnan2025,
  title = {Conditional Mutual Information and Information-Theoretic Phases of Decohered Gibbs States},
  author = {Zhang, Yifan F. and Gopalakrishnan, Sarang},
  journal = {Phys. Rev. Lett.},
  volume = {135},
  issue = {16},
  pages = {160401},
  numpages = {6},
  year = {2025},
  month = {Oct},
  publisher = {American Physical Society},
  doi = {10.1103/cjkh-1z6c},
  url = {https://link.aps.org/doi/10.1103/cjkh-1z6c}
}

@article{Shaebani_2020,
   title={Computational models for active matter},
   volume={2},
   ISSN={2522-5820},
   url={http://dx.doi.org/10.1038/s42254-020-0152-1},
   DOI={10.1038/s42254-020-0152-1},
   number={4},
   journal={Nature Reviews Physics},
   publisher={Springer Science and Business Media LLC},
   author={Shaebani, M. Reza and Wysocki, Adam and Winkler, Roland G. and Gompper, Gerhard and Rieger, Heiko},
   year={2020},
   month=mar, pages={181–199} }

@article{Greenacre2022PCA,
  author  = {Greenacre, Michael and Primicerio, Raul and others},
  title   = {Principal component analysis},
  journal = {Nature Reviews Methods Primers},
  year    = {2022},
  volume  = {2},
  pages   = {100},
  doi     = {10.1038/s43586-022-00184-w}
}

@article{TroppWebber2023RandomizedLowRank,
  title={Randomized algorithms for low-rank matrix approximation: Design, analysis, and applications},
  author={Tropp, Joel A and Webber, Robert J},
  journal={arXiv preprint arXiv:2306.12418},
  year={2023},
  url={https://arxiv.org/abs/2306.12418}
}

@article{Ajanki_2018,
   title={Stability of the matrix Dyson equation and random matrices with correlations},
   volume={173},
   ISSN={1432-2064},
   url={http://dx.doi.org/10.1007/s00440-018-0835-z},
   DOI={10.1007/s00440-018-0835-z},
   number={1–2},
   journal={Probability Theory and Related Fields},
   publisher={Springer Science and Business Media LLC},
   author={Ajanki, Oskari H. and Erdős, László and Krüger, Torben},
   year={2018},
   month=feb, pages={293–373} }

@misc{mariella2025orderparameterdiscoveryquantum,
  title         = {Order Parameter Discovery for Quantum Many-Body Systems},
  author        = {Mariella, Nicola and Murphy, Tara and Di Marcantonio, Francesco and Najafi, Khadijeh and Vallecorsa, Sofia and Zhuk, Sergiy and Rico, Enrique},
  year          = {2025},
  eprint        = {2408.01400},
  archivePrefix = {arXiv},
  primaryClass  = {quant-ph},
  url           = {https://arxiv.org/abs/2408.01400}
}

@article{Shi_2015,
   title={Universal Order Parameters and Quantum Phase Transitions: A Finite-Size Approach},
   volume={5},
   ISSN={2045-2322},
   url={http://dx.doi.org/10.1038/srep07673},
   DOI={10.1038/srep07673},
   number={1},
   journal={Scientific Reports},
   publisher={Springer Science and Business Media LLC},
   author={Shi, Qian-Qian and Zhou, Huan-Qiang and Batchelor, Murray T.},
   year={2015},
   month=jan }

@article{Evers_2008,
   title={Anderson transitions},
   volume={80},
   ISSN={1539-0756},
   url={http://dx.doi.org/10.1103/RevModPhys.80.1355},
   DOI={10.1103/revmodphys.80.1355},
   number={4},
   journal={Reviews of Modern Physics},
   publisher={American Physical Society (APS)},
   author={Evers, Ferdinand and Mirlin, Alexander D.},
   year={2008},
   month=oct, pages={1355–1417} }

@article{shklovskii,
  title = {Statistics of spectra of disordered systems near the metal-insulator transition},
  author = {Shklovskii, B. I. and Shapiro, B. and Sears, B. R. and Lambrianides, P. and Shore, H. B.},
  journal = {Phys. Rev. B},
  volume = {47},
  issue = {17},
  pages = {11487--11490},
  numpages = {0},
  year = {1993},
  month = {May},
  publisher = {American Physical Society},
  doi = {10.1103/PhysRevB.47.11487},
  url = {https://link.aps.org/doi/10.1103/PhysRevB.47.11487}
}

@article{Mirlin2000,
   title={Statistics of energy levels and eigenfunctions in disordered systems},
   volume={326},
   ISSN={0370-1573},
   url={http://dx.doi.org/10.1016/S0370-1573(99)00091-5},
   DOI={10.1016/s0370-1573(99)00091-5},
   number={5-6},
   journal={Physics Reports},
   publisher={Elsevier BV},
   author={Mirlin, A},
   year={2000},
   month=Mar, pages={259–382} }

@misc{saberi2026longrangecorrelatedrandommatrices,
  author        = {Saberi, Abbas Ali and Moessner, Roderich},
  title         = {Long-Range Correlated Random Matrices},
  year          = {2026},
  eprint        = {2604.22447v2},
  archivePrefix = {arXiv},
  primaryClass  = {cond-mat.stat-mech}
}

@book{Cardy_1996, place={Cambridge}, series={Cambridge Lecture Notes in Physics}, title={Scaling and Renormalization in Statistical Physics}, publisher={Cambridge University Press}, author={Cardy, John}, year={1996}, collection={Cambridge Lecture Notes in Physics}}

@article{cicuta2018unifying,
  title={Unifying model for random matrix theory in arbitrary space dimensions},
  author={Cicuta, Giovanni M and Krausser, Johannes and Milkus, Rico and Zaccone, Alessio},
  journal={Physical Review E},
  volume={97},
  number={3},
  pages={032113},
  year={2018},
  publisher={APS}
}
\bibliographystyle{apsrev4-2}
\onecolumngrid

\begin{appendix}
\section{Correlation Structure Statistical Physics Random Matrices in Fourier Space\label{sec:corrcalculation}}
In this section, we derive the correlation structure of the Fourier-transformed random matrix ensemble. 
\begin{align}
    \langle \hat{\mathcal{M}}_{k_x k_y}\hat{\mathcal{M}}_{q_x q_y}\rangle 
    &= \frac{1}{L^2}\left\langle\sum_{x,y,\tilde{x},\tilde{y}=0}^{L-1}
    e^{\frac{2\pi i}{L} \left( (k_x x - k_y y) + (q_x \tilde{x} - q_y \tilde{y}) \right)}
    \mathcal{M}_{xy}\mathcal{M}_{\tilde{x}\tilde{y}} \right\rangle, \\ 
    &= \frac{1}{L^3}\sum_{x,y,\tilde{x},\tilde{y}=0}^{L-1} 
    e^{\frac{2\pi i}{L} \left( (k_x x - k_y y) + (q_x \tilde{x} - q_y \tilde{y}) \right)}
    (G(x - \tilde{x}, y - \tilde{y})+G(x-\tilde{y},y-\tilde{x}))\label{eqn:momcorr}
\end{align}
where $G(x,y)$ is the translation-invariant 2-point correlator of the parameter field, and all differences are taken modulo $L$. By an appropriate change of variables, eq.~\eqref{eqn:momcorr} can be written as  
\begin{align}
   \langle \hat{\mathcal{M}}_{k_x k_y}\hat{\mathcal{M}}_{q_x q_y}\rangle 
   &= \left(\frac{1}{L^3} \sum_{\Delta x, \Delta y\in \mathbb{Z}_L}
   \sum_{\tilde{x}, \tilde{y}=0}^{L-1}
   e^{\frac{2\pi i}{L} \left(
   (k_x+q_x)\tilde{x} - (k_y+q_y)\tilde{y}
   + k_x \Delta x - k_y \Delta y \right)}
   G(\Delta x, \Delta y) \right)+\left((q_x,q_y)\longleftrightarrow (-q_y,-q_x)\right) \\
   &= \frac{1}{L}(\delta_{k_x,q_y}\delta_{q_x,k_y}+\delta_{k_x, -q_x} \delta_{k_y, -q_y}) \,
   \sum_{\Delta x, \Delta y\in \mathbb{Z}_L}
   e^{\frac{2\pi i}{L} (k_x \Delta x - k_y \Delta y)}
   G(\Delta x, \Delta y)  \\
   &=  (\delta_{k_x,q_y} \delta_{k_y ,q_x}+\delta_{k_x,-q_x}\delta_{k_y,-q_y})\mathcal{V}_{k_xk_y} \label{eqn:corrstructure}
\end{align}
where the \textit{matrix of variances} is defined as
\begin{equation}
\mathcal{V}_{k_xk_y}
:= \frac{1}{L}\sum_{\Delta x\in \mathbb{Z}_L} \sum_{\Delta y\in \mathbb{Z}_L}
e^{\frac{2\pi i}{L} (k_x \Delta x - k_y \Delta y)}
G(\Delta x, \Delta y).
\end{equation}
The model thus behaves like a \textit{Wigner-type matrix} (see~\cite{ajanki2017universalitygeneralwignertypematrices}) subject to an additional dependence arising from the second term in eq.~\eqref{eqn:corrstructure}. 

\end{appendix}
\end{document}